\documentclass[twocolumn,amssymb,aps,showpacs,longbibliography]{revtex4-1}
\usepackage{graphicx}
\usepackage[colorlinks]{hyperref}
\usepackage{amsbsy}

\begin{document}

\title{Influence of Landau level mixing on the properties of \\
elementary excitations in graphene in strong magnetic field}

\author{Yu.\,E. Lozovik${}^{1,2}$}\email{lozovik@isan.troitsk.ru}
\author{A.\,A. Sokolik${}^1$}

\affiliation{${}^1$Institute for Spectroscopy RAS, 142190, Troitsk, Moscow Region, Russia\\
${}^2$Moscow Institute of Physics and Technology, 141700, Moscow, Russia}%

\begin{abstract}
Massless Dirac electrons in graphene fill Landau levels with energies scaled as square roots of their numbers. Coulomb
interaction between electrons leads to mixing of different Landau levels. The relative strength of this interaction
depends only on dielectric susceptibility of surrounding medium and can be large in suspended graphene. We consider
influence of Landau level mixing on the properties of magnetoexcitons and magnetoplasmons --- elementary electron-hole
excitations in graphene in quantizing magnetic field. We show that, at small enough background dielectric screening,
the mixing leads to very essential change of magnetoexciton and magnetoplasmon dispersion laws in comparison with the
lowest Landau level approximation.
\end{abstract}


\maketitle

\section*{Introduction}

Two-dimensional systems in strong magnetic field are studied intensively since the discovery of integer and fractional
quantum Hall effects \cite{Klitzing,Stormer,DasSarma}. For a long time, such systems were represented by gallium
arsenide heterostructures with two-dimensional electron motion within each subband \cite{Ando}.

New and very interesting realization of two-dimensional electron system appeared when graphene, a monoatomic layer of
carbon, was successfully isolated \cite{NovoselovScience,NovoselovNature}. The most spectacular property of graphene is
the fact that its electrons behave as massless chiral particles, obeying Dirac equation. Intensive experimental and
theoretical studies of this material over several recent years yielded a plethora of interesting results
\cite{GrapheneRMP,SokolikUFN,Abergel}. In particular, graphene demonstrates unusual half-integer quantum Hall effect
\cite{NovoselovNature}, which can be observed even at room temperature \cite{NovoselovQHE}.

In external perpendicular magnetic field, the motion of electrons along cyclotron orbits acquires zero-dimensional
character and, as a result, electrons fill discrete Landau levels \cite{Landau}. In semiconductor quantum wells, Landau
levels are equidistant and separation between them is determined by the cyclotron frequency $\omega_\mathrm{c}=eH/mc$.
In graphene, due to massless nature of electrons, ``ultra-relativistic'' Landau levels appear, which are
non-equidistant and located symmetrically astride the Dirac point \cite{Zheng,Gusynin}. Energies of these levels are
$E_n=\mathrm{sign}(n)\sqrt{2|n|}v_\mathrm{F}/l_H$, where $n=0,\pm1,\pm2,\ldots$, $v_\mathrm{F}\approx10^6\,\mbox{m/s}$
is the Fermi velocity of electrons and $l_H=\sqrt{c/eH}$ is magnetic length, or radius of the cyclotron orbit (here and
below we assume $\hbar=1$).

In the case of integer filling, when several Landau levels are completely filled by electrons and all higher levels are
empty, elementary excitations in the system are caused by electron transitions from one of the filled Landau levels to
one of the empty levels \cite{KallinHalperin}. Such transitions can be observed in cyclotron resonance or Raman
scattering experiments as absorption peaks at ceratin energies. With neglect of Coulomb interaction, energy of the
excited electron-hole pair is just a distance between Landau levels of electron and hole. Coulomb interaction leads to
mixing of transitions between different pairs of Landau levels, changing the resulting energies of elementary
excitations.

Characteristic energy of Coulomb interaction in magnetic field is $e^2/\varepsilon l_H$, where $\varepsilon$ is a
dielectric permittivity of surrounding medium. The relative strength of Coulomb interaction can be estimated as ratio
of its characteristic value to a characteristic distance between Landau levels. For massive electrons in semiconductor
quantum wells, this ratio is proportional to $H^{-1/2}$, thus in asymptotically strong magnetic field the Coulomb
interaction becomes a weak perturbation \cite{Lai,Lerner1}. In this case, the lowest Landau level approximation,
neglecting Landau level mixing, is often used. It was shown that Bose-condensate of noninteracting magnetoexcitons in
the lowest Landau level as an exact ground state in semiconductor quantum well in strong magnetic field
\cite{DzyubenkoJPA}.

A different situation arises in graphene. The relative strength of Coulomb interaction in this system can be expressed
as $r_\mathrm{s}=e^2/\varepsilon v_\mathrm{F}$ and does not depend on magnetic field \cite{Goerbig2}. The only
parameter which can influence it is the dielectric permittivity of surrounding medium $\varepsilon$. At small enough
$\varepsilon$, mixing between different Landau levels can significantly change properties of elementary excitations in
graphene.

Coulomb interaction leads to appearance of two types of elementary excitations from the filled Landau levels. From
summation of ``ladder'' diagrams we get magnetoexcitons, which can be imagined as bound states of electron and hole in
magnetic field \cite{KallinHalperin,Lerner1,Lerner2}. Properties of magnetoexcitons in graphene were considered in
several works, mainly in the lowest Landau level approximation \cite{Iyengar,Bychkov,SokolikPSSA,Koinov,Roldan1}. At
$\varepsilon\approx3$, Landau level mixing was shown to be weak in the works \cite{Iyengar,Zhang}.

Note that influence of Landau level mixing on properties of an insulating ground state of neutral graphene was
considered in \cite{Jung} by means of tight-binding Hartree-Fock approximation. It was shown that Landau level mixing
favors formation of insulating charge-density wave state instead of ferromagnetic and spin-density wave states in
suspended graphene, i.e. at weak enough background dielectric screening.

From the experimental point of view, the most interesting are magnetoexcitons with zero total momentum, which are only
able to couple with electromagnetic radiation due to very small photon momentum. For usual non-relativistic electrons,
magnetoexciton energy at zero momentum is protected against corrections due to electron interactions by the Kohn
theorem \cite{Kohn}. However, for electrons with linear dispersion in graphene the Kohn theorem is not applicable
\cite{Jiang,Bychkov,Roldan1,Henriksen,Shizuya2,Orlita,Pyatkovskiy}. Thus, observable energies of excitonic spectral
lines can be seriously renormalized relatively to the bare values, calculated without taking into account Coulomb
interaction.

The other type of excitations can be derived using the random phase approximation, corresponding to summation of
``bubble'' diagrams. These excitations, called magnetoplasmons, are analogue of plasmons and have been studied both in
two-dimensional electron gas \cite{Chiu,KallinHalperin} and graphene
\cite{Shizuya1,Tahir1,Berman2,Bychkov,Roldan1,Iyengar,Roldan3,Roldan4,Goerbig2,Fischer2} (both with and without taking
into account Landau level mixing).

In the present article, we consider magnetoexcitons and magnetoplasmons with taking into account Landau level mixing
and show how the properties of these excitations change in comparison with the lowest Landau level approximation. For
magnetoexcitons, we take into account the mixing of asymptotically large number of Landau levels and find the limiting
values of cyclotron resonance energies.

For simplicity and in order to stress the role of virtual transitions between different pairs of electron and hole
Landau levels (i.e. the role of two-particle processes), here we do not take into account renormalization of
single-particle energies via exchange with the filled levels. This issue have been studied in several theoretical works
\cite{Barlas,Bychkov,Iyengar,Roldan1,Shizuya2}. Correction of Landau level energies can be treated as renormalization
of the Fermi velocity, dependent on the ultraviolet cutoff for a number of the filled Landau levels taken into account
in exchange processes.

The rest of this article is organized as follows. In Section 2 we present a formalism for description of
magnetoexcitons in graphene, which is applied in Section 3 to study influence of Coulomb interaction and Landau level
mixing on their properties. In Section 4 we study magnetoplasmons in graphene in the random phase approximation and in
Section 5 we formulate the conclusions.

\section{Magnetoexcitons}

Electrons in graphene populate vicinities of two nonequivalent Dirac points in the Brillouin zone, or two valleys
$\mathbf{K}$ and $\mathbf{K}'$. We do not consider intervalley scattering and neglect valley splitting, thus it is
sufficient to consider electrons in only one valley and treat existence of the other valley as additional twofold
degeneracy of electron states.

We consider magnetoexciton as an electron-hole pair, and we will denote all electron and hole variables by the indices
1 and 2 respectively. In the valley $\mathbf{K}$, Hamiltonian of free electrons in graphene in the basis $\{A_1,B_2\}$
of sublattices takes a form \cite{GrapheneRMP}:
\begin{eqnarray}
H_1^{(0)}=v_\mathrm{F}\sqrt2\left(\begin{array}{cc}0&p_{1-}\\p_{1+}&0\end{array}\right),\label{H10}
\end{eqnarray}
where $p_{1\pm}=(p_{1x}\pm ip_{1y})/\sqrt2$ are the cyclic components of electron momentum and
$v_\mathrm{F}\approx10^6\,\mbox{m/s}$ is the Fermi velocity of electrons.

For external magnetic field $\mathbf{H}$, parallel to the $z$ axis, we take the symmetrical gauge, when
$\mathbf{A}(\mathbf{r})=\frac12[\mathbf{H}\times\mathbf{r}]$. Introducing the magnetic field as substitution of the
momentum $\mathbf{p}_1\rightarrow\mathbf{p}_1+(e/c)\mathbf{A}(\mathbf{r}_1)$ in (\ref{H10}) (we treat the electron
charge as $-e$), we get the Hamiltonian of the form:
\begin{eqnarray}
H_1=\frac{v_\mathrm{F}\sqrt2}{l_H}\left(\begin{array}{cc}0&a_1\\a^+_1&0\end{array}\right).\label{H1}
\end{eqnarray}
Here the operators $a_1=l_Hp_{1-}-ir_{1-}/2l_H$ and $a_1^+=l_Hp_{1+}+ir_{1+}/2l_H$ (where $r_{1\pm}=(x_1\pm
iy_1)/\sqrt2$) obey bosonic commutation relation $[a_1,a^+_1]=1$.

Using this relation, by means of successive action of the raising operator $a_1^+$ we can construct Landau levels for
electron \cite{Goerbig2} with energies
\begin{eqnarray}
E^\mathrm{L}_n=s_n\sqrt{2|n|}\frac{v_\mathrm{F}}{l_H}\label{E_L}
\end{eqnarray}
and wave functions
\begin{eqnarray}
\psi_{nk}(\mathbf{r})=\left(\sqrt2\right)^{\delta_{n0}-1}\left(\begin{array}{c}s_n\phi_{|n|-1,k}(\mathbf{r})\\
\phi_{|n|k}(\mathbf{r})\end{array}\right).
\end{eqnarray}
Here $k=0,1,2,\ldots$ is the index of guiding center, which enumerates electron states on the $n$-th Landau level
($n=-\infty,\ldots,+\infty$), having macroscopically large degeneracy $N_\phi=S/2\pi l_H^2$, equal to a number of
magnetic flux quanta penetrating the system of the area $S$. Eigenfunctions $\phi_{nk}(\mathbf{r})$ of a
two-dimensional harmonic oscillator have the explicit form:
\begin{eqnarray}
\phi_{nk}(\mathbf{r})=\frac{i^{|n-k|}}{\sqrt{2\pi}l_H}\sqrt{\frac{\min(n,k)!}{\max(n,k)!}}\,e^{-r^2/4l_H^2}\nonumber\\
\times\left(\frac{x+is_{n-k}y}{\sqrt2l_H}\right)^{|n-k|}L_{\min(n,k)}^{|n-k|}\left(\frac{r^2}{2l_H^2}\right),
\end{eqnarray}
$s_n=\mathrm{sign}(n)$ and $L_n^\alpha(x)$ are associated Laguerre polynomials.

Consider now the hole states. A hole wave function is a complex conjugated electron wave function, and the hole charge
is $+e$. Thus, we can obtain Hamiltonian of the hole in magnetic field from the electron Hamiltonian (\ref{H1}) by
complex conjugation and reversal of the sign of the vector potential $\mathbf{A}(\mathbf{r}_2)$. In the representation
of sublattices $\{A_2,B_2\}$ it is
\begin{eqnarray}
H_2=\frac{v_\mathrm{F}\sqrt2}{l_H}\left(\begin{array}{cc}0&a_2\\a^+_2&0\end{array}\right),\label{H2}
\end{eqnarray}
where the operators $a_2=l_Hp_{2+}-ir_{2+}/2l_H$ and $a_2^+=l_Hp_{2-}+ir_{2-}/2l_H$ commute with $a_1$, $a_1^+$ and
obey the commutation relation $[a_2,a_2^+]=1$. Energies of the hole Landau levels are the same as these of electron
Landau levels (\ref{E_L}), but have an opposite sign.

Hamiltonian of electron-hole pair without taking into account Landau level mixing is just the sum of (\ref{H1}) and
(\ref{H2}), and can be represented in the combined basis of electron and hole sublattices
$\{A_1A_2,A_1B_2,B_1A_2,B_1B_2\}$ as
\begin{eqnarray}
H_0=H_1+H_2=\frac{v_\mathrm{F}\sqrt2}{l_H}\left(\begin{array}{cccc}0&a_2&a_1&0\\a_2^+&0&0&a_1\\
a_1^+&0&0&a_2\\0&a_1^+&a_2^+&0\end{array}\right).\label{H0}
\end{eqnarray}

It is known \cite{Gor'kov} that for electron-hole pair in magnetic field there exists a conserving two-dimensional
vector of magnetic momentum, equal in our gauge to
\begin{eqnarray}
\mathbf{P}=\mathbf{p}_1+\mathbf{p}_2-\frac{e}{2c}[\mathbf{H}\times(\mathbf{r}_1-\mathbf{r}_2)]\label{P}
\end{eqnarray}
and playing the role of a center-of-mass momentum. The magnetic momentum is a generator of simultaneous translation in
space and gauge transformation, preserving invariance of Hamiltonian of charged particles in magnetic field
\cite{Ruvinsky}.

The magnetic momentum commutes with both the noninteracting Hamiltonian $(\ref{H0})$ and electron-hole Coulomb
interaction $V(\mathbf{r}_1-\mathbf{r}_2)$. Therefore, we can find a wave function of magnetoexciton as an
eigenfunction of (\ref{P}):
\begin{eqnarray}
\Psi_{\mathbf{P}n_1n_2}(\mathbf{r}_1,\mathbf{r}_2)=\frac1{2\pi}
\exp\left\{i\mathbf{R}\left(\mathbf{P}+\frac{[\mathbf{e}_z\times\mathbf{r}]}{2l_H^2}\right)\right\}\nonumber\\
\times\Phi_{n_1n_2}(\mathbf{r}-\mathbf{r}_0).\label{Psi}
\end{eqnarray}
Here $\mathbf{R}=(\mathbf{r}_1+\mathbf{r}_2)/2$, $\mathbf{r}=\mathbf{r}_1-\mathbf{r}_2$, $\mathbf{e}_z$ is a unit
vector in the direction of the $z$ axis. The wave function of relative motion of electron and hole
$\Phi_{n_1n_2}(\mathbf{r}-\mathbf{r}_0)$ is shifted on the vector $\mathbf{r}_0=l_H^2[\mathbf{e}_z\times\mathbf{P}]$.
This shift can be attributed to electric field, appearing in the moving reference frame of magnetoexciton and pulling
apart electron and hole.

Transformation (\ref{Psi}) from $\Psi$ to $\Phi$ can be considered as a unitary transformation $\Phi=U\Psi$,
corresponding a to a switching from the laboratory reference frame to the magnetoexciton rest frame. Accordingly, we
should transform operators as $A\rightarrow SAS^+$. Transforming the operators in (\ref{H0}), we get: $Sa_1S^+=b_1$,
$Sa_1^+S^+=b_1^+$, $Sa_2S^+=-b_2$, $Sb_2^+S^+=-b_2^+$. Here the operators $b_1=l_Hp_--ir_-/2l_H$,
$b_1^+=l_Hp_++ir_+/2l_H$, $b_2=l_Hp_+-ir_+/2l_H$, $b_2^+=l_Hp_-+ir_-/2l_H$ contain only the relative electron-hole
coordinate and momentum and obey commutation relations $[b_1,b_1^+]=1$, $[b_2,b_2^+]=1$ (all other commutators vanish).

Thus, the Hamiltonian (\ref{H0}) of electron-hole pair in its center-of-mass reference frame takes the form
\begin{eqnarray}
H_0'=\frac{v_\mathrm{F}\sqrt2}{l_H}\left(\begin{array}{cccc}0&-b_2&b_1&0\\-b_2^+&0&0&b_1\\
b_1^+&0&0&-b_2\\0&b_1^+&-b_2^+&0\end{array}\right).\label{H0rel}
\end{eqnarray}
A four-component wave function of electron-hole relative motion $\Phi_{n_1n_2}$, being an eigenfunction of
(\ref{H0rel}), can be constructed by successive action of the raising operators $b_1^+$ and $b_2^+$ (see also
\cite{Iyengar,Bychkov}):
\begin{eqnarray}
\Phi_{n_1n_2}(\mathbf{r})=\left(\sqrt2\right)^{\delta_{n_1,0}+\delta_{n_2,0}-2}\nonumber\\
\times\left(\begin{array}{c}
s_{n_1}s_{n_2}\phi_{|n_1|-1,|n_2|-1}(\mathbf{r})\\s_{n_1}\phi_{|n_1|-1,|n_2|}(\mathbf{r})\\
s_{n_2}\phi_{|n_1|,|n_2|-1}(\mathbf{r})\\ \phi_{|n_1||n_2|}(\mathbf{r})\end{array}\right).\label{Phi}
\end{eqnarray}
The bare energy of magnetoexciton in this state is a difference between energies (\ref{E_L}) of electron and hole
Landau levels:
\begin{eqnarray}
E_{n_1n_2}^{(0)}=E_{n_1}^\mathrm{L}-E_{n_2}^\mathrm{L}.\label{E0}
\end{eqnarray}

Here we label the state of relative motion by numbers of Landau levels $n_1$ and $n_2$ of electron and hole
respectively. The whole wave function of magnetoexciton (\ref{Psi}) is additionally labeled by the magnetic momentum
$\mathbf{P}$. In the case of integer filling, when all Landau levels up to $\nu$-th one are completely filled by
electrons and all upper levels are empty, magnetoexciton states with $n_1>\nu$, $n_2\leq\nu$ are possible. For
simplicity, we neglect Zeeman and valley splittings of electron states, leading to appearance of additional spin-flip
and intervalley excitations \cite{Iyengar,Bychkov,Roldan1}.

\section{Influence of Coulomb interaction}

Now we take into account the Coulomb interaction between electron and hole $V(\mathbf{r})=-e^2/\varepsilon r$, screened
by surrounding dielectric medium with permittivity $\varepsilon$. Upon switching into the electron-hole center-of-mass
reference frame, it is transformed as $V'(\mathbf{r})=V(\mathbf{r}+\mathbf{r}_0)$. To obtain magnetoexciton energies
with taking into account Coulomb interaction, we should find eigenvalues of the full Hamiltonian of relative motion
$H'=H_0'+V'$ in the basis of the bare magnetoexcitonic states (\ref{Phi}). As discussed in the Introduction, a relative
strength of the Coulomb interaction is described by the dimensionless parameter
\begin{eqnarray}
r_\mathrm{s}=\frac{e^2}{\varepsilon v_\mathrm{F}}\approx\frac{2.2}\varepsilon.
\end{eqnarray}

\begin{figure*}[!t]
\includegraphics[width=0.9\textwidth]{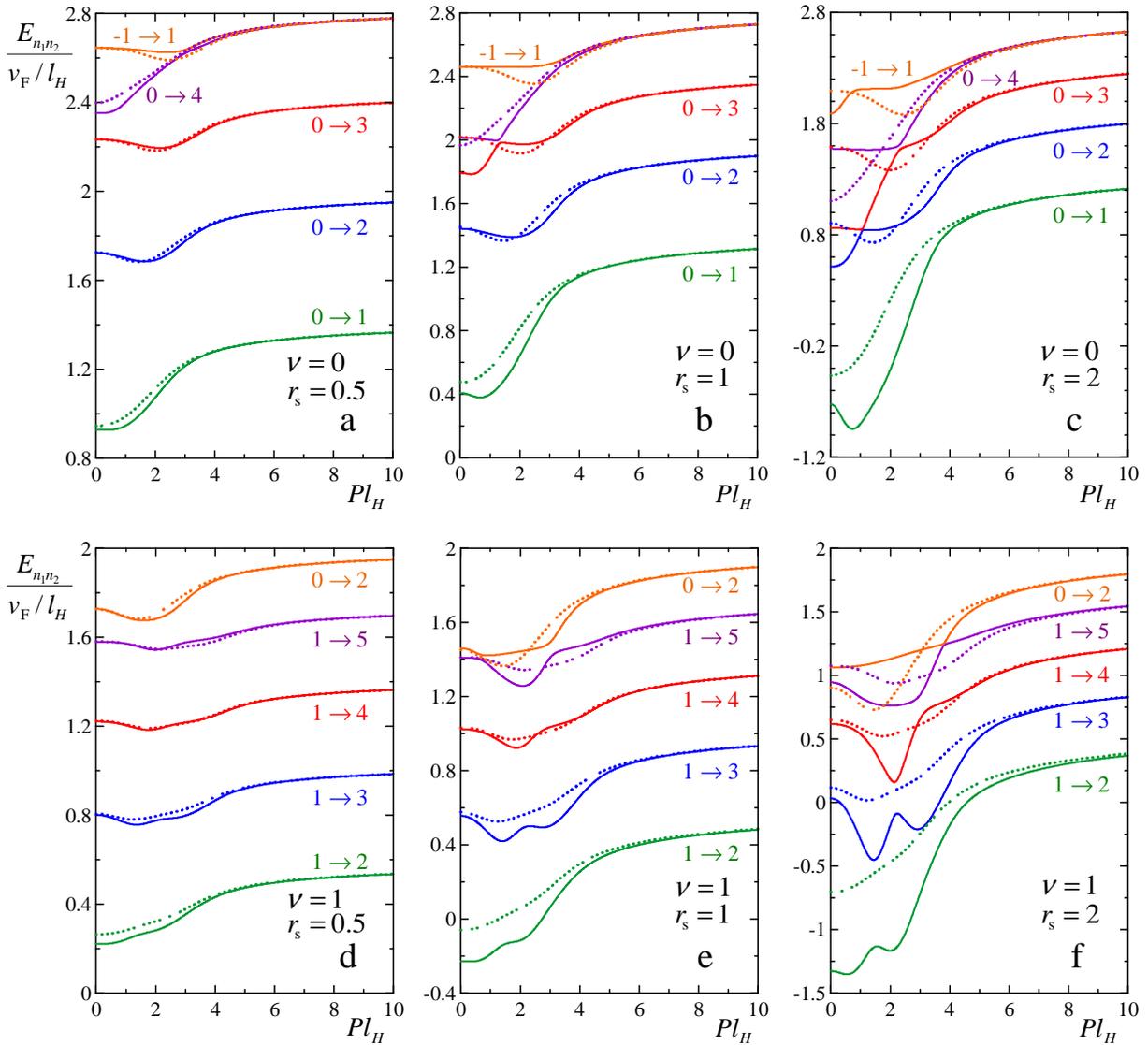}
\caption{Magnetoexciton dispersions $E_{n_1n_2}(P)$, calculated in the first order in Coulomb interaction (dotted
lines) and with taking into account mixing between 16 low-lying magnetoexciton states (solid lines). The dispersions
are calculated at different filling factors $\nu$ and different $r_\mathrm{s}$: (a) $\nu=0$, $r_\mathrm{s}=0.5$, (b)
$\nu=0$, $r_\mathrm{s}=1$, (c) $\nu=0$, $r_\mathrm{s}=2$, (d) $\nu=1$, $r_\mathrm{s}=0.5$, (e) $\nu=1$,
$r_\mathrm{s}=1$, (f) $\nu=1$, $r_\mathrm{s}=2$. Dispersions of 5 lowest-lying magnetoexciton states $n_2\rightarrow
n_1$, indicated near the corresponding curves, are shown.}\label{Fig1}
\end{figure*}

When $\varepsilon\gg1$, $r_\mathrm{s}\ll1$ and we can treat Coulomb interaction as a weak perturbation and calculate
magnetoexciton energy in the first order in the interaction as:
\begin{eqnarray}
E_{n_1n_2}^{(1)}(P)=E_{n_1n_2}^{(0)}+\langle\Phi_{n_1n_2}|V'|\Phi_{n_1n_2}\rangle.\label{E1}
\end{eqnarray}
Due to spinor nature of electron wave functions in graphene, the correction (\ref{E1}) to the bare magnetoexciton
energy (\ref{E0}) is a sum of four terms, each of them having a form of correction to magnetoexciton energy in usual
two-dimensional electron gas \cite{Iyengar,Bychkov,SokolikPSSA}. Dependence of magnetoexciton energy on magnetic
momentum $\mathbf{P}$ can be attributed to Coulomb interaction between electron and hole, separated by the average
distance $r_0\propto P$.

Calculations of magnetoexciton dispersions in the first order in Coulomb interaction (\ref{E1}) have been performed in
several works \cite{Bychkov,Iyengar,Roldan1,SokolikPSSA,Koinov}. However, such calculations are well-justified only at
small enough $r_\mathrm{s}$, i.e. at large $\varepsilon$. When $\varepsilon\sim1$ (this is achievable in experiments
with suspended graphene \cite{Bolotin,Ghahari,Elias,Knox}), the role of virtual electron transitions between different
Landau levels can be significant.

\begin{figure*}[!t]
\includegraphics[width=0.9\textwidth]{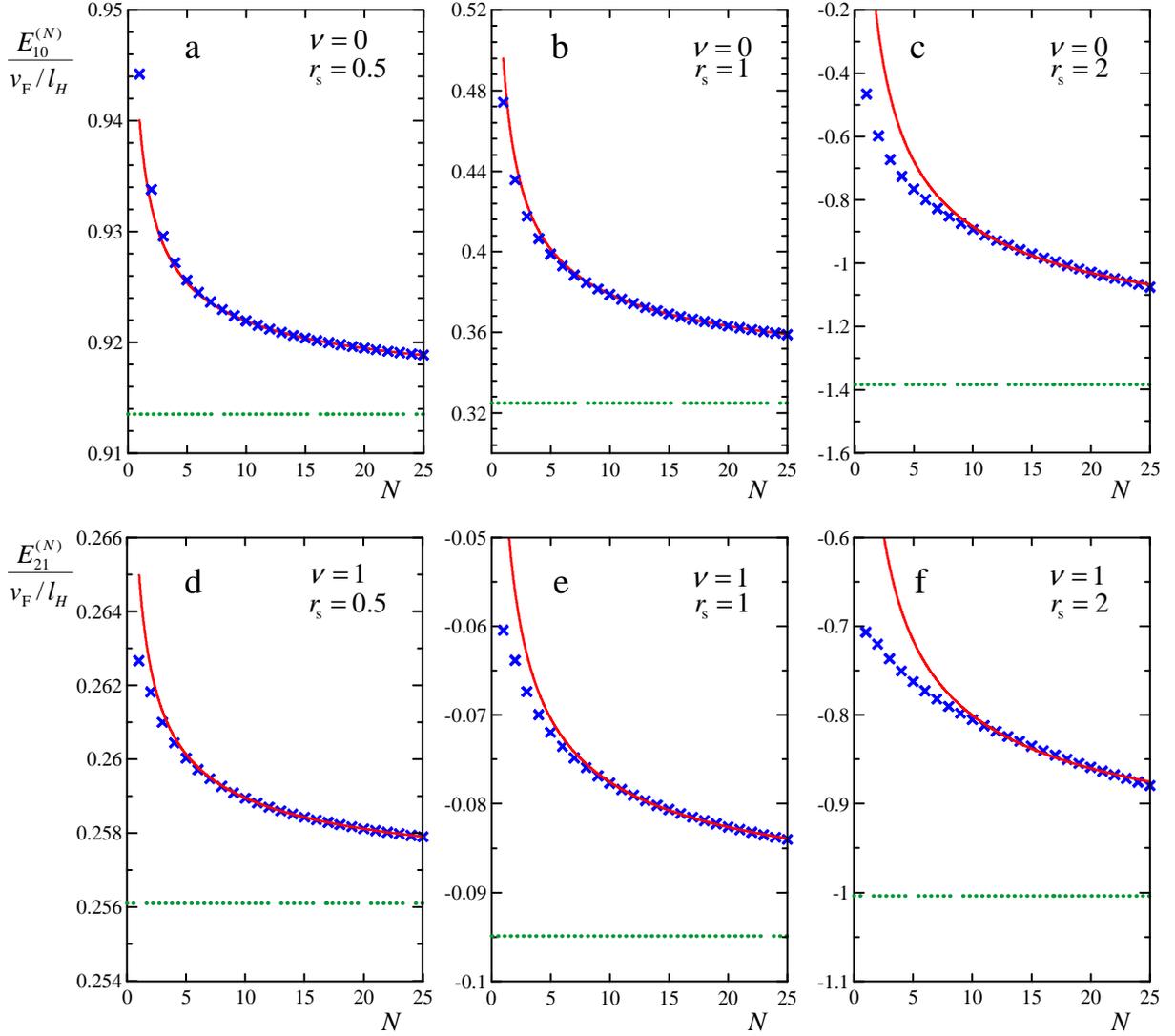}
\caption{Magnetoexciton energies at rest $E_{n_1n_2}^{(N)}(P=0)$, calculated with taking into account $N$ electron and
$N$ hole Landau levels, with stepwise increasing $N$ (crosses). The fits to these energies with inverse-square-root
function (solid lines) and limiting values of $E_{n_1n_2}^{(N)}(P=0)$ at $N\rightarrow\infty$ (dotted lines) are also
shown. The results are presented for different filling factors $\nu$ and different $r_\mathrm{s}$: (a) $\nu=0$,
$r_\mathrm{s}=0.5$, (b) $\nu=0$, $r_\mathrm{s}=1$, (c) $\nu=0$, $r_\mathrm{s}=2$, (d) $\nu=1$, $r_\mathrm{s}=0.5$, (e)
$\nu=1$, $r_\mathrm{s}=1$, (f) $\nu=1$, $r_\mathrm{s}=2$.}\label{Fig2}
\end{figure*}

To take into account Landau level mixing, we should perform diagonalization of full Hamiltonian of Coulomb interacting
electrons in some basis of magnetoexcitonic states $\Psi_{\mathbf{P}n_1n_2}$, where electron Landau levels $n_1>\nu$
are unoccupied and hole Landau levels $n_2\leq\nu$ are occupied. To obtain eigenvalues of the Hamiltonian, we need to
solve the equation:
\begin{eqnarray}
\det\left\|\delta_{n_1'n_1}\delta_{n_2'n_2}(E_{n_1n_2}^{(0)}-E)\right.\nonumber\\
\left.+\langle\Psi_{\mathbf{P}n_1'n_2'}|V|\Psi_{\mathbf{P}n_1'n_2'}\rangle\vphantom{E_{n_1n_2}^{(0)}}\right\|=0.
\label{diag}
\end{eqnarray}
We can constrain our basis to $N^2$ terms, involving $N$ Landau levels for electron ($n_1=\nu+1,\ldots,\nu+N$) and $N$
Landau levels for a hole ($n_2=\nu,\ldots,\nu-N+1$). Since the Hamiltonian commutes with magnetic momentum
$\mathbf{P}$, the procedure of diagonalization can be performed independently at different values of $\mathbf{P}$,
resulting in dispersions $E_{n_1n_2}^{(N)}(P)$ of magnetoexcitons, affected by a mixing between $N$ electron and $N$
hole Landau levels.

We present in Fig.~\ref{Fig1} dispersion relations for 5 lowest magnetoexciton states, calculated with and without
taking into account the mixing between 16 lowest-energy states. The results are shown for Landau level fillings $\nu=0$
and $\nu=1$, and for different values of $r_\mathrm{s}$. Close to $P=0$, magnetoexciton can be described as a composite
particle with parabolic dispersion, characterized by some effective mass
$M_{n_1n_2}=[d^2E_{n_1n_2}(P)/dP^2]^{-1}|_{P=0}$. At large $P$, the Coulomb interaction weakens and the dispersions
tend to the energies of one-particle excitations (\ref{E0}). However, the dispersion can be rather complicated
structure with several minima and maxima at intermediate momenta $P\sim l_H^{-1}$.

We see that the mixing at small $r_\mathrm{s}$ has a weak effect on the dispersions (solid and dotted lines are very
close in Fig.~\ref{Fig1}(a,d)). However, at $r_\mathrm{s}\sim1$ the mixing changes the dispersions significantly. We
can observe avoided crossings between dispersions of different magnetoexcitons, and even reversal of a sign of
magnetoexciton effective masses (see Fig.~\ref{Fig1}(b,c,e,f)). Also we see that the high levels are more strongly
mixed than the low-lying ones. Similar results were presented in \cite{Iyengar} for $r_\mathrm{s}=0.73$ with conclusion
that the mixing is weak.

As we see, at large $r_\mathrm{s}$ the mixing of several Landau levels already strongly changes magnetoexciton
dispersions. Important question arises here: how many Landau levels should we take into account to achieve convergency
of results? To answer this question, we perform diagonalization of the type (\ref{diag}), increasing step-by-step a
quantity $N$ of electron and hole Landau levels. For simplicity, we perform these calculations at $P=0$ only. Energies
of magnetoexcitons at rest, renormalized by electron interactions due to breakdown of the Kohn theorem, are the most
suitable to be observed in optical experiments.

The results of such calculations of $E_{n_1n_2}^{(N)}(P=0)$ as functions of $N$ are shown in Fig.~\ref{Fig2} by cross
points. We found semi-analytically that eigenvalues of the Hamiltonian under consideration should approach a dependence
\begin{eqnarray}
E_{n_1n_2}^{(N)}\approx E_{n_1n_2}^{(\infty)}+\frac{C_{n_1n_2}}{\sqrt{N}}
\end{eqnarray}
at large $N$. We fitted the numerical results by this dependence and thus were able to find the limiting values
$E_{n_1n_2}^{(\infty)}$ of magnetoexciton energies with infinite number of Landau levels taken into account.

We see in Fig.~\ref{Fig2} that the differences between magnetoexciton energies calculated in the first order in Coulomb
interaction (the crosses at $N=1$) and the energies calculated with taking into account mixing between all Landau
levels (dotted lines) are very small at $r_\mathrm{s}=0.5$ (Fig.~\ref{Fig2}(a,b)), moderate at $r_\mathrm{s}=1$
(Fig.~\ref{Fig2}(b,e)) and very large at $r_\mathrm{s}=2$ (Fig.~\ref{Fig2}(c,f)). Since convergency of the
inverse-square-root function is very slow, even the mixing of rather large (of the order of tens) number of Landau
levels is not sufficient to obtain reliable results for magnetoexciton energies, as clearly seen in the
Fig.~\ref{Fig2}.

Note that the mixing increases magnetoexciton binding energies, similarly to results on magnetoexcitons in
semiconductor quantum wells \cite{Moskalenko1,Moskalenko2}.

\section{Magnetoplasmons}

Magnetoplasmons are collective excitations of electron gas in magnetic field, occurring as poles of density-to-density
response function. In the random phase approximation, dispersion of magnetoplasmon is determined as a root of the
equation
\begin{eqnarray}
1-V(q)\Pi(q,\omega)=0,\label{RPA}
\end{eqnarray}
where $V(q)=2\pi e^2/\varepsilon q$ is the two-dimensional Fourier transform of Coulomb interaction and $\Pi(q,\omega)$
is a polarization operator (or polarizability). Polarization operator for graphene in magnetic field can be expressed
using magnetoexciton wave functions (\ref{Phi}) and energies (\ref{E0}) (see also
\cite{Shizuya1,Tahir1,Berman2,Pyatkovskiy,Roldan3,Roldan4,Goerbig2}):
\begin{eqnarray}
\Pi(q,\omega)=g\sum_{n_1n_2}\frac{f_{n_2}-f_{n1}}{\omega-E_{n_1n_2}^{(0)}+i\delta}F_{n_1n_2}(q),\label{Pol}
\end{eqnarray}
\begin{eqnarray}
F_{n_1n_2}(q)=\Phi^+_{n_1n_2}(ql_H^2)\nonumber\\
\times\left(\begin{array}{cccc}1&0&0&1\\0&0&0&0\\0&0&0&0\\1&0&0&1\end{array}\right) \Phi_{n_1n_2}(ql_H^2),\label{F}
\end{eqnarray}
where $g=4$ is the degeneracy factor and $f_n$ is the occupation number for the $n$-th Landau level, i.e. $f_n=1$ at
$n\leq\nu$ and $f_n=0$ at $n>\nu$ (we neglect temperature effects since typical separation between Landau levels in
graphene in quantizing magnetic field is of the order of room temperature \cite{NovoselovQHE}). The matrix between
magnetoexcitonic wave functions in (\ref{F}) ensures that electron and hole belong to the same sublattice, that is
needed for Coulomb interaction in exchange channel treated as annihilation of electron and hole in one point of space
and subsequent creation of electron-hole pair in another point.

Unlike electron gas without magnetic field, having a single plasmon branch, Eqs.~(\ref{RPA})--(\ref{F}) give an
infinite number of solutions $\omega=\Omega_{n_1n_2}(q)$, each of them can be attributed to specific inter-Landau level
transition $n_2\rightarrow n_1$, affected by Coulomb interaction \cite{Roldan3,Roldan4,Goerbig2}. Note that at
$q\rightarrow0$, when Coulomb interaction $V(q)$ becomes weak, dispersion of each magnetoplasmon branch
$\Omega_{n_1n_2}(q)$ tends to the corresponding single-particle excitation energy $E_{n_1n_2}^{(0)}$.

At $r_\mathrm{s}\ll1$, we can suppose that magnetoplasmon energy $\Omega_{n_1n_2}(q)$ does not differ significantly
from the single-particle energy $E_{n_1n_2}^{(0)}$. In this case a dominant contribution to the sum in (\ref{Pol})
comes from the term with the given $n_1$ and $n_2$. Neglecting all other terms, we can write (\ref{Pol}) as
\begin{eqnarray}
\Pi(q,\omega)\approx g\frac{F_{n_1n_2}(q)}{\omega-E_{n_1n_2}^{(0)}+i\delta},\label{Pol1}
\end{eqnarray}
and from (\ref{RPA}) we obtain an approximation to plasmon dispersion in the first order in the Coulomb interaction:
\begin{eqnarray}
\Omega_{n_1n_2}(q)\approx E_{n_1n_2}^{(0)}+gV(q)F_{n_1n_2}(q).\label{Omega}
\end{eqnarray}

\begin{figure*}[!t]
\includegraphics[width=0.9\textwidth]{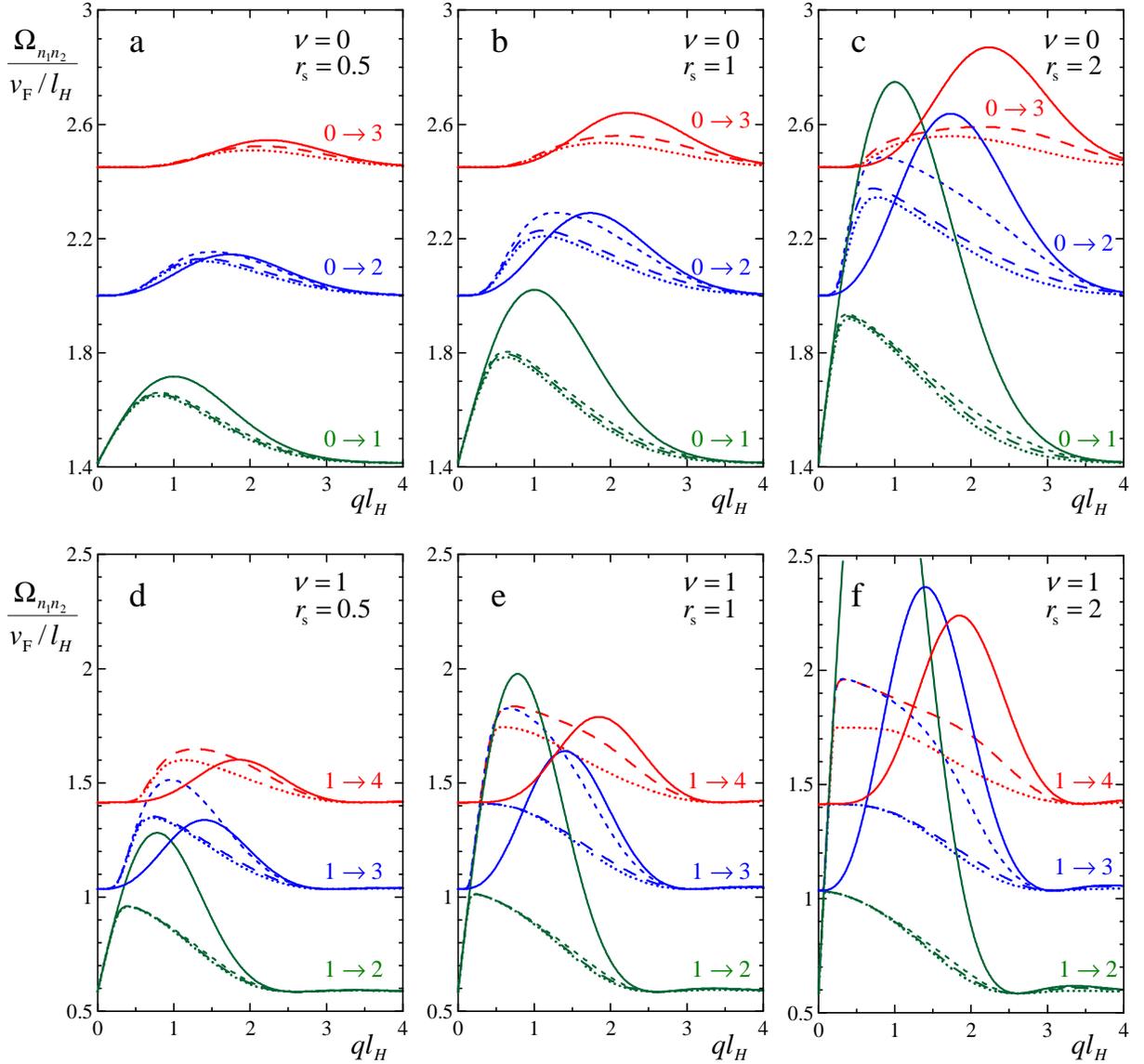}
\caption{Magnetoplasmon energies $\Omega_{n_1n_2}$, calculated in the lowest Landau level approximation (solid lines),
with taking into account mixing between 2 (short dash lines) and 3 (long dash lines) Landau levels of electron and
hole, and with taking into account mixing between all Landau levels (dotted lines). The results are presented for
different filling factors $\nu$ and different $r_\mathrm{s}$: (a) $\nu=0$, $r_\mathrm{s}=0.5$, (b) $\nu=0$,
$r_\mathrm{s}=1$, (c) $\nu=0$, $r_\mathrm{s}=2$, (d) $\nu=1$, $r_\mathrm{s}=0.5$, (e) $\nu=1$, $r_\mathrm{s}=1$, (f)
$\nu=1$, $r_\mathrm{s}=2$. Dispersions of 3 lowest-lying magnetoplasmon modes $n_2\rightarrow n_1$, indicated near the
corresponding curves, are shown.}\label{Fig3}
\end{figure*}

Magnetoplasmons in graphene were considered without taking into account Landau level mixing in a manner of
Eq.~(\ref{Omega}) in the works \cite{Iyengar,Fischer2}. Other authors \cite{Shizuya1,Bychkov,Roldan1} took into account
several Landau levels, and the others \cite{Berman2,Tahir1,Roldan3,Roldan4} performed full summation in the framework
of the random phase approximation (\ref{RPA})--(\ref{F}) to calculate magnetoplasmon dispersions.

Here we state the question: how many Landau levels one should take into account to calculate magnetoplasmon spectrum
with sufficient accuracy? To answer it, we performed calculations with successive taking into account increasing number
of Landau levels at different $\nu$ and $r_\mathrm{s}$. In Fig.~\ref{Fig3}, dispersions of magnetoplasmons in graphene
calculated numerically are shown. Results obtained without taking into account Landau level mixing, with taking into
account a mixing of two or three lowest Landau levels and with taking into account all Landau levels are plotted with
different line styles.

As we see, even taking into account the mixing between 2 Landau levels changes the dispersions considerably (see the
differences between solid and short dash lines in Fig.~\ref{Fig3}). However, the calculations with mixing between 3
Landau levels (long dash lines) are already close to the exact results (dotted lines), except for the high-lying
magnetoplasmon modes. It is also seen, that the mixing considerably changes the dispersions even at moderate
$r_\mathrm{s}$ (see, e.g., Fig.~\ref{Fig3}(d) at $r_\mathrm{s}=0.5$). Note that the mixing usually decrease
magnetoplasmon energies and does not affect the long-wavelength linear asymptotics of their dispersions.

Therefore we conclude here that convergency of magnetoplasmon dispersions in rather fast upon increasing a number of
Landau levels taken into account. Several lowest Landau levels are sufficient to obtain rather accurate results. On the
other hand, calculations in the lowest Landau level approximation, i.e. without taking into account the mixing, can
give inaccurate results, especially in a region of intermediate momenta $q\sim l_H^{-1}$.

\section{Conclusions}

We studied influence of Landau level mixing in graphene in quantizing magnetic field on properties of elementary
excitations --- magnetoexcitons and magnetoplasmons --- in this system. Virtual transitions between Landau levels,
caused by Coulomb interaction, can change dispersions of the excitations in comparison with the lowest Landau level
approximation.

Strength of Coulomb interaction and thus a degree of Landau level mixing can be characterized by dimensionless
parameter $r_\mathrm{s}$, dependent in the case of graphene only on dielectric permittivity of surrounding medium. By
embedding graphene in different environments, one can change $r_\mathrm{s}$ from small values to $r_\mathrm{s}\approx2$
\cite{Morozov}.

We calculated dispersions of magnetoexcitons in graphene and showed that the mixing even between few Landau levels can
change significantly the dispersion curves at $r_\mathrm{s}>1$. However, at small $r_\mathrm{s}$ the role of the mixing
is negligible, in agreement with the other works \cite{Iyengar,Zhang}. Then the question about convergency of such
calculations upon increasing a number of involved Landau levels have been raised.

We performed calculations of magnetoexciton energies at rest with taking into account stepwise increasing number of
Landau levels and found their inverse-square-root asymptotics. By evaluating limiting values of these asymptotics, we
calculated magnetoexciton energies with infinite number of Landau levels taken into account. We demonstrated that
influence of remote Landau levels of magnetoexciton energies is strong, especially at large $r_\mathrm{s}$. Also it was
found that calculations with taking into account even several Landau levels provide results, rather far from exact
ones.

Also dispersion relations of magnetoplasmons in graphene were calculated in the random phase approximation with taking
into account different numbers of Landau levels. We showed that even few Landau levels for electron and hole are
sufficient do obtain accurate results, however the lowest Landau level approximation (i.e. calculations without taking
into account the mixing) provide inaccurate results, especially for intermediate momenta and high-lying magnetoplasmon
modes.

In out study, we completely disregarded renormalization of single-particle energies due to exchange with filled Landau
levels in the valence band of graphene, since this question was considered elsewhere
\cite{Barlas,Bychkov,Iyengar,Roldan1,Shizuya2}. In our article, we focused on the role of Coulomb interaction in the
electron-hole channel only. In this context, an important result of our work is that breakdown of the Kohn theorem in
graphene leads to strong corrections of magnetoexciton energies not only due to exchange self-energies, but also due to
virtual transitions caused by Coulomb interaction between electron and hole.

We considered magnetoexcitons in the ladder approximation and magnetoplasmons in the random phase approximations
without taking into account vertex corrections and screening. Estimating the role of these factors, especially in the
strong-interacting regime at large $r_\mathrm{s}$, is a difficult task and will be postponed for future studies.

The results obtained in this work should be relevant for magneto-optical spectroscopy of graphene
\cite{Orlita,Jiang,Henriksen,Sadowski,Kashuba,Giesbers} and for the problem of Bose-condensation of magnetoexcitons
\cite{Berman1,Fil,Bezuglyi}. Excitonic lines in optical absorption or Raman spectra of graphene can give experimental
information about energies of elementary excitations. Magnetoexcitons and magnetoplasmons can be observed also as
constituents of various hybrid modes --- polaritons \cite{Berman3}, trions \cite{Fischer1}, Bernstein modes
\cite{Roldan2} or magnetophonon resonances \cite{Goerbig1}.

The work was supported by grants of Russian Foundation for Basic Research and by the grant of the President of Russian
Federation for Young Scientists MK-5288.2011.2. One of the authors (A.A.S.) also acknowledges support from the Dynasty
Foundation.

\bibliography{article}

\begin{thebibliography}{10}%
\makeatletter
\providecommand \@ifxundefined [1]{%
 \ifx #1\undefined \expandafter \@firstoftwo
 \else \expandafter \@secondoftwo
\fi
}%
\providecommand \@ifnum [1]{%
 \ifnum #1\expandafter \@firstoftwo
 \else \expandafter \@secondoftwo
\fi
}%
\providecommand \enquote [1]{``#1''}%
\providecommand \bibnamefont  [1]{#1}%
\providecommand \bibfnamefont [1]{#1}%
\providecommand \citenamefont [1]{#1}%
\providecommand\href[0]{\@sanitize\@href}%
\providecommand\@href[1]{\endgroup\@@startlink{#1}\endgroup\@@href}%
\providecommand\@@href[1]{#1\@@endlink}%
\providecommand \@sanitize [0]{\begingroup\catcode`\&12\catcode`\#12\relax}%
\@ifxundefined \pdfoutput {\@firstoftwo}{%
 \@ifnum{\z@=\pdfoutput}{\@firstoftwo}{\@secondoftwo}%
}{%
 \providecommand\@@startlink[1]{\leavevmode\special{html:<a href="#1">}}%
 \providecommand\@@endlink[0]{\special{html:</a>}}%
}{%
 \providecommand\@@startlink[1]{%
  \leavevmode
  \pdfstartlink
   attr{/Border[0 0 1 ]/H/I/C[0 1 1]}%
   user{/Subtype/Link/A<</Type/Action/S/URI/URI(#1)>>}%
  \relax
 }%
 \providecommand\@@endlink[0]{\pdfendlink}%
}%
\providecommand \url  [0]{\begingroup\@sanitize \@url }%
\providecommand \@url [1]{\endgroup\@href {#1}{\urlprefix}}%
\providecommand \urlprefix [0]{URL }%
\providecommand \Eprint[0]{\href }%
\@ifxundefined \urlstyle {%
  \providecommand \doi [1]{doi:\discretionary{}{}{}#1}%
}{%
  \providecommand \doi [0]{doi:\discretionary{}{}{}\begingroup
  \urlstyle{rm}\Url }%
}%
\providecommand \doibase [0]{http://dx.doi.org/}%
\providecommand \Doi[1]{\href{\doibase#1}}%
\providecommand \bibAnnote [3]{%
  \BibitemShut{#1}%
  \begin{quotation}\noindent
    \textsc{Key:}\ #2\\\textsc{Annotation:}\ #3%
  \end{quotation}%
}%
\providecommand \bibAnnoteFile [2]{%
  \IfFileExists{#2}{\bibAnnote {#1} {#2} {\input{#2}}}{}%
}%
\providecommand \typeout [0]{\immediate \write \m@ne }%
\providecommand \selectlanguage [0]{\@gobble}%
\providecommand \bibinfo [0]{\@secondoftwo}%
\providecommand \bibfield [0]{\@secondoftwo}%
\providecommand \translation [1]{[#1]}%
\providecommand \BibitemOpen[0]{}%
\providecommand \bibitemStop [0]{}%
\providecommand \bibitemNoStop [0]{.\EOS\space}%
\providecommand \EOS [0]{\spacefactor3000\relax}%
\providecommand \BibitemShut [1]{\csname bibitem#1\endcsname}%
\bibitem{Klitzing}%
  \BibitemOpen
  \bibfield{author}{%
  \bibinfo {author} {\bibfnamefont{K}~\bibnamefont{von Klitzing}},\ }%
  \bibfield{title}{%
  \enquote{\bibinfo {title} {The quantized hall effect},}\ }%
  \bibfield{journal}{%
  \bibinfo {journal} {Rev. Mod. Phys.}\ }%
  \textbf{\bibinfo {volume} {58}},\ \bibinfo {pages} {519--531} (\bibinfo
  {year} {1986})%
  \bibAnnoteFile{NoStop}{Klitzing}%
\bibitem{Stormer}%
  \BibitemOpen
  \bibfield{author}{%
  \bibinfo {author} {\bibfnamefont{H~L}\ \bibnamefont{Stormer}}, \bibinfo
  {author} {\bibfnamefont{D~C}\ \bibnamefont{Tsui}},\ and\ \bibinfo {author}
  {\bibfnamefont{A~C}\ \bibnamefont{Gossard}},\ }%
  \bibfield{title}{%
  \enquote{\bibinfo {title} {The fractional quantum hall effect},}\ }%
  \bibfield{journal}{%
  \bibinfo {journal} {Rev. Mod. Phys.}\ }%
  \textbf{\bibinfo {volume} {71}},\ \bibinfo {pages} {S298--S305} (\bibinfo
  {year} {1999})%
  \bibAnnoteFile{NoStop}{Stormer}%
\bibitem{DasSarma}%
  \BibitemOpen
  \bibfield{author}{%
  \bibinfo {author} {\bibfnamefont{S}~\bibnamefont{Das~Sarma}}\ and\ \bibinfo
  {author} {\bibfnamefont{A}~\bibnamefont{Pinczuk}},\ }%
  \emph{\bibinfo {title} {Perspectives in quantum Hall effects}}\ (\bibinfo
  {publisher} {Wiley},\ \bibinfo {address} {New York},\ \bibinfo {year}
  {1997})%
  \bibAnnoteFile{NoStop}{DasSarma}%
\bibitem{Ando}%
  \BibitemOpen
  \bibfield{author}{%
  \bibinfo {author} {\bibfnamefont{T}~\bibnamefont{Ando}}, \bibinfo {author}
  {\bibfnamefont{A~B}\ \bibnamefont{Fowler}},\ and\ \bibinfo {author}
  {\bibfnamefont{F}~\bibnamefont{Stern}},\ }%
  \bibfield{title}{%
  \enquote{\bibinfo {title} {Electronic properties of two-dimensional
  systems},}\ }%
  \bibfield{journal}{%
  \bibinfo {journal} {Rev. Mod. Phys.}\ }%
  \textbf{\bibinfo {volume} {54}},\ \bibinfo {pages} {437--672} (\bibinfo
  {year} {1981})%
  \bibAnnoteFile{NoStop}{Ando}%
\bibitem{NovoselovScience}%
  \BibitemOpen
  \bibfield{author}{%
  \bibinfo {author} {\bibfnamefont{K~S}\ \bibnamefont{Novoselov}}, \bibinfo
  {author} {\bibfnamefont{A~K}\ \bibnamefont{Geim}}, \bibinfo {author}
  {\bibfnamefont{S~V}\ \bibnamefont{Morozov}}, \bibinfo {author}
  {\bibfnamefont{D}~\bibnamefont{Jiang}}, \bibinfo {author}
  {\bibfnamefont{Y}~\bibnamefont{Zhang}}, \bibinfo {author}
  {\bibfnamefont{S~V}\ \bibnamefont{Dubonos}}, \bibinfo {author}
  {\bibfnamefont{I~V}\ \bibnamefont{Grigorieva}},\ and\ \bibinfo {author}
  {\bibfnamefont{A~A}\ \bibnamefont{Firsov}},\ }%
  \bibfield{title}{%
  \enquote{\bibinfo {title} {Electric field effect in atomically thin carbon
  films},}\ }%
  \bibfield{journal}{%
  \bibinfo {journal} {Science}\ }%
  \textbf{\bibinfo {volume} {306}},\ \bibinfo {pages} {666--669} (\bibinfo
  {year} {2004})%
  \bibAnnoteFile{NoStop}{NovoselovScience}%
\bibitem{NovoselovNature}%
  \BibitemOpen
  \bibfield{author}{%
  \bibinfo {author} {\bibfnamefont{K~S}\ \bibnamefont{Novoselov}}, \bibinfo
  {author} {\bibfnamefont{A~K}\ \bibnamefont{Geim}}, \bibinfo {author}
  {\bibfnamefont{S~V}\ \bibnamefont{Morozov}}, \bibinfo {author}
  {\bibfnamefont{D}~\bibnamefont{Jiang}}, \bibinfo {author}
  {\bibfnamefont{M~I}\ \bibnamefont{Katsnelson}}, \bibinfo {author}
  {\bibfnamefont{I~V}\ \bibnamefont{Grigorieva}}, \bibinfo {author}
  {\bibfnamefont{S~V}\ \bibnamefont{Dubonos}},\ and\ \bibinfo {author}
  {\bibfnamefont{A~A}\ \bibnamefont{Firsov}},\ }%
  \bibfield{title}{%
  \enquote{\bibinfo {title} {Two-dimensional gas of massless dirac fermions in
  graphene},}\ }%
  \bibfield{journal}{%
  \bibinfo {journal} {Nature}\ }%
  \textbf{\bibinfo {volume} {438}},\ \bibinfo {pages} {197--200} (\bibinfo
  {year} {2005})%
  \bibAnnoteFile{NoStop}{NovoselovNature}%
\bibitem{GrapheneRMP}%
  \BibitemOpen
  \bibfield{author}{%
  \bibinfo {author} {\bibfnamefont{A~H}\ \bibnamefont{Castro~Neto}}, \bibinfo
  {author} {\bibfnamefont{F}~\bibnamefont{Guinea}}, \bibinfo {author}
  {\bibfnamefont{N~M~R}\ \bibnamefont{Peres}}, \bibinfo {author}
  {\bibfnamefont{K~S}\ \bibnamefont{Novoselov}},\ and\ \bibinfo {author}
  {\bibfnamefont{A~K}\ \bibnamefont{Geim}},\ }%
  \bibfield{title}{%
  \enquote{\bibinfo {title} {The electronic properties of graphene},}\ }%
  \bibfield{journal}{%
  \bibinfo {journal} {Rev. Mod. Phys.}\ }%
  \textbf{\bibinfo {volume} {80}},\ \bibinfo {pages} {109--162} (\bibinfo
  {year} {2009})%
  \bibAnnoteFile{NoStop}{GrapheneRMP}%
\bibitem{SokolikUFN}%
  \BibitemOpen
  \bibfield{author}{%
  \bibinfo {author} {\bibfnamefont{Yu~E}\ \bibnamefont{Lozovik}}, \bibinfo
  {author} {\bibfnamefont{S~P}\ \bibnamefont{Merkulova}},\ and\ \bibinfo
  {author} {\bibfnamefont{A~A}\ \bibnamefont{Sokolik}},\ }%
  \bibfield{title}{%
  \enquote{\bibinfo {title} {Collective electron phenomena in graphene},}\ }%
  \bibfield{journal}{%
  \bibinfo {journal} {Phys.-Usp.}\ }%
  \textbf{\bibinfo {volume} {51}},\ \bibinfo {pages} {727--748} (\bibinfo
  {year} {2008})%
  \bibAnnoteFile{NoStop}{SokolikUFN}%
\bibitem{Abergel}%
  \BibitemOpen
  \bibfield{author}{%
  \bibinfo {author} {\bibfnamefont{D~S~L}\ \bibnamefont{Abergel}}, \bibinfo
  {author} {\bibfnamefont{V}~\bibnamefont{Apalkov}}, \bibinfo {author}
  {\bibfnamefont{J}~\bibnamefont{Berashevich}}, \bibinfo {author}
  {\bibfnamefont{K}~\bibnamefont{Ziegler}},\ and\ \bibinfo {author}
  {\bibfnamefont{T}~\bibnamefont{Chakraborty}},\ }%
  \bibfield{title}{%
  \enquote{\bibinfo {title} {Properties of graphene: A theoretical
  perspective},}\ }%
  \bibfield{journal}{%
  \bibinfo {journal} {Adv. in Phys.}\ }%
  \textbf{\bibinfo {volume} {59}},\ \bibinfo {pages} {261--482} (\bibinfo
  {year} {2010})%
  \bibAnnoteFile{NoStop}{Abergel}%
\bibitem{NovoselovQHE}%
  \BibitemOpen
  \bibfield{author}{%
  \bibinfo {author} {\bibfnamefont{K~S}\ \bibnamefont{Novoselov}}, \bibinfo
  {author} {\bibfnamefont{Z}~\bibnamefont{Jiang}}, \bibinfo {author}
  {\bibfnamefont{Y}~\bibnamefont{Zhang}}, \bibinfo {author}
  {\bibfnamefont{S~V}\ \bibnamefont{Morozov}}, \bibinfo {author}
  {\bibfnamefont{H~L}\ \bibnamefont{Stormer}}, \bibinfo {author}
  {\bibfnamefont{U}~\bibnamefont{Zeitler}}, \bibinfo {author}
  {\bibfnamefont{G~S}\ \bibnamefont{Maan}, \bibfnamefont{J~Cand~Boebinger}},
  \bibinfo {author} {\bibfnamefont{P}~\bibnamefont{Kim}},\ and\ \bibinfo
  {author} {\bibfnamefont{A~K}\ \bibnamefont{Geim}},\ }%
  \bibfield{title}{%
  \enquote{\bibinfo {title} {Room-temperature quantum hall effect in
  graphene},}\ }%
  \bibfield{journal}{%
  \bibinfo {journal} {Science}\ }%
  \textbf{\bibinfo {volume} {315}},\ \bibinfo {pages} {1379--1379} (\bibinfo
  {year} {2007})%
  \bibAnnoteFile{NoStop}{NovoselovQHE}%
\bibitem{Landau}%
  \BibitemOpen
  \bibfield{author}{%
  \bibinfo {author} {\bibfnamefont{L~D}\ \bibnamefont{Landau}}\ and\ \bibinfo
  {author} {\bibfnamefont{E~M}\ \bibnamefont{Lifshitz}},\ }%
  \emph{\bibinfo {title} {Quantum mechanics, 3rd edition}}\ (\bibinfo
  {publisher} {Butterworth-Heinemann},\ \bibinfo {address} {New York},\
  \bibinfo {year} {1981})%
  \bibAnnoteFile{NoStop}{Landau}%
\bibitem{Zheng}%
  \BibitemOpen
  \bibfield{author}{%
  \bibinfo {author} {\bibfnamefont{Y}~\bibnamefont{Zheng}}\ and\ \bibinfo
  {author} {\bibfnamefont{T}~\bibnamefont{Ando}},\ }%
  \bibfield{title}{%
  \enquote{\bibinfo {title} {Hall conductivity of a two-dimensional graphite
  system},}\ }%
  \bibfield{journal}{%
  \bibinfo {journal} {Phys. Rev. B}\ }%
  \textbf{\bibinfo {volume} {65}},\ \bibinfo {pages} {245420} (\bibinfo {year}
  {2002})%
  \bibAnnoteFile{NoStop}{Zheng}%
\bibitem{Gusynin}%
  \BibitemOpen
  \bibfield{author}{%
  \bibinfo {author} {\bibfnamefont{V~P}\ \bibnamefont{Gusynin}}\ and\ \bibinfo
  {author} {\bibfnamefont{S~G}\ \bibnamefont{Sharapov}},\ }%
  \bibfield{title}{%
  \enquote{\bibinfo {title} {Unconventional integer quantum hall effect in
  graphene},}\ }%
  \bibfield{journal}{%
  \bibinfo {journal} {Phys. Rev. Lett.}\ }%
  \textbf{\bibinfo {volume} {95}},\ \bibinfo {pages} {146801} (\bibinfo {year}
  {2005})%
  \bibAnnoteFile{NoStop}{Gusynin}%
\bibitem{KallinHalperin}%
  \BibitemOpen
  \bibfield{author}{%
  \bibinfo {author} {\bibfnamefont{C}~\bibnamefont{Kallin}}\ and\ \bibinfo
  {author} {\bibfnamefont{B~I}\ \bibnamefont{Halperin}},\ }%
  \bibfield{title}{%
  \enquote{\bibinfo {title} {Excitations from a filled landau level in the
  two-dimensional electron gas},}\ }%
  \bibfield{journal}{%
  \bibinfo {journal} {Phys. Rev. B}\ }%
  \textbf{\bibinfo {volume} {30}},\ \bibinfo {pages} {5655--5668} (\bibinfo
  {year} {1984})%
  \bibAnnoteFile{NoStop}{KallinHalperin}%
\bibitem{Lai}%
  \BibitemOpen
  \bibfield{author}{%
  \bibinfo {author} {\bibfnamefont{D}~\bibnamefont{Lai}},\ }%
  \bibfield{title}{%
  \enquote{\bibinfo {title} {Matter in strong magnetic fields},}\ }%
  \bibfield{journal}{%
  \bibinfo {journal} {Rev. Mod. Phys,}\ }%
  \textbf{\bibinfo {volume} {73}},\ \bibinfo {pages} {629--661} (\bibinfo
  {year} {2001})%
  \bibAnnoteFile{NoStop}{Lai}%
\bibitem{Lerner1}%
  \BibitemOpen
  \bibfield{author}{%
  \bibinfo {author} {\bibfnamefont{I~V}\ \bibnamefont{Lerner}}\ and\ \bibinfo
  {author} {\bibfnamefont{Yu~E}\ \bibnamefont{Lozovik}},\ }%
  \bibfield{title}{%
  \enquote{\bibinfo {title} {Mott exciton in a quasi--two-dimensional
  semiconductor in a strong magnetic field},}\ }%
  \bibfield{journal}{%
  \bibinfo {journal} {Sov. Phys. JETP}\ }%
  \textbf{\bibinfo {volume} {51}},\ \bibinfo {pages} {588--592} (\bibinfo
  {year} {1980})%
  \bibAnnoteFile{NoStop}{Lerner1}%
\bibitem{DzyubenkoJPA}%
  \BibitemOpen
  \bibfield{author}{%
  \bibinfo {author} {\bibfnamefont{A~B}\ \bibnamefont{Dzyubenko}}\ and\
  \bibinfo {author} {\bibfnamefont{Yu~E}\ \bibnamefont{Lozovik}},\ }%
  \bibfield{title}{%
  \enquote{\bibinfo {title} {Symmetry of hamiltonians of quantum two-component
  systems: condensate of composite particles as an exact eigenstate},}\ }%
  \bibfield{journal}{%
  \bibinfo {journal} {J. Phys. A: Math. Gen.}\ }%
  \textbf{\bibinfo {volume} {24}},\ \bibinfo {pages} {415--424} (\bibinfo
  {year} {1991})%
  \bibAnnoteFile{NoStop}{DzyubenkoJPA}%
\bibitem{Goerbig2}%
  \BibitemOpen
  \bibfield{author}{%
  \bibinfo {author} {\bibfnamefont{M~O}\ \bibnamefont{Goerbig}},\ }%
  \bibfield{title}{%
  \enquote{\bibinfo {title} {Electronic properties of graphene in a strong
  magnetic field},}\ }%
  \bibfield{journal}{%
  \bibinfo {journal} {Rev. Mod. Phys.}\ }%
  \textbf{\bibinfo {volume} {83}},\ \bibinfo {pages} {1193--1243} (\bibinfo
  {year} {2011})%
  \bibAnnoteFile{NoStop}{Goerbig2}%
\bibitem{Lerner2}%
  \BibitemOpen
  \bibfield{author}{%
  \bibinfo {author} {\bibfnamefont{I~V}\ \bibnamefont{Lerner}}\ and\ \bibinfo
  {author} {\bibfnamefont{Yu~E}\ \bibnamefont{Lozovik}},\ }%
  \bibfield{title}{%
  \enquote{\bibinfo {title} {Two-dimensional electron-hole system in a strong
  magnetic field as an almost ideal exciton gas},}\ }%
  \bibfield{journal}{%
  \bibinfo {journal} {Sov. Phys. JETP}\ }%
  \textbf{\bibinfo {volume} {53}},\ \bibinfo {pages} {763--770} (\bibinfo
  {year} {1981})%
  \bibAnnoteFile{NoStop}{Lerner2}%
\bibitem{Iyengar}%
  \BibitemOpen
  \bibfield{author}{%
  \bibinfo {author} {\bibfnamefont{A}~\bibnamefont{Iyengar}}, \bibinfo {author}
  {\bibfnamefont{J}~\bibnamefont{Wang}}, \bibinfo {author} {\bibfnamefont{H~A}\
  \bibnamefont{Fertig}},\ and\ \bibinfo {author}
  {\bibfnamefont{L}~\bibnamefont{Brey}},\ }%
  \bibfield{title}{%
  \enquote{\bibinfo {title} {Excitations from filled landau levels in
  graphene},}\ }%
  \bibfield{journal}{%
  \bibinfo {journal} {Phys. Rev. B}\ }%
  \textbf{\bibinfo {volume} {75}},\ \bibinfo {pages} {125430} (\bibinfo {year}
  {2007})%
  \bibAnnoteFile{NoStop}{Iyengar}%
\bibitem{Bychkov}%
  \BibitemOpen
  \bibfield{author}{%
  \bibinfo {author} {\bibfnamefont{Yu~A}\ \bibnamefont{Bychkov}}\ and\ \bibinfo
  {author} {\bibfnamefont{G}~\bibnamefont{Martinez}},\ }%
  \bibfield{title}{%
  \enquote{\bibinfo {title} {Magnetoplasmon excitations in graphene for filling
  factors $\nu\leq6$},}\ }%
  \bibfield{journal}{%
  \bibinfo {journal} {Phys. Rev. B}\ }%
  \textbf{\bibinfo {volume} {77}},\ \bibinfo {pages} {125417} (\bibinfo {year}
  {2008})%
  \bibAnnoteFile{NoStop}{Bychkov}%
\bibitem{SokolikPSSA}%
  \BibitemOpen
  \bibfield{author}{%
  \bibinfo {author} {\bibfnamefont{Yu~E}\ \bibnamefont{Lozovik}}, \bibinfo
  {author} {\bibfnamefont{A~A}\ \bibnamefont{Sokolik}},\ and\ \bibinfo {author}
  {\bibfnamefont{M}~\bibnamefont{Willander}},\ }%
  \bibfield{title}{%
  \enquote{\bibinfo {title} {Collective phases and magnetoexcitons in
  graphene},}\ }%
  \bibfield{journal}{%
  \bibinfo {journal} {Phys. Stat. Sol. A}\ }%
  \textbf{\bibinfo {volume} {206}},\ \bibinfo {pages} {927--930} (\bibinfo
  {year} {2009})%
  \bibAnnoteFile{NoStop}{SokolikPSSA}%
\bibitem{Koinov}%
  \BibitemOpen
  \bibfield{author}{%
  \bibinfo {author} {\bibfnamefont{Z~G}\ \bibnamefont{Koinov}},\ }%
  \bibfield{title}{%
  \enquote{\bibinfo {title} {Magnetoexciton dispersion in graphene bilayers
  embedded in a dielectric},}\ }%
  \bibfield{journal}{%
  \bibinfo {journal} {Phys. Rev. B}\ }%
  \textbf{\bibinfo {volume} {79}},\ \bibinfo {pages} {073409} (\bibinfo {year}
  {2009})%
  \bibAnnoteFile{NoStop}{Koinov}%
\bibitem{Roldan1}%
  \BibitemOpen
  \bibfield{author}{%
  \bibinfo {author} {\bibfnamefont{R}~\bibnamefont{Rold\'{a}n}}, \bibinfo
  {author} {\bibfnamefont{J-N}\ \bibnamefont{Fuchs}},\ and\ \bibinfo {author}
  {\bibfnamefont{M~O}\ \bibnamefont{Goerbig}},\ }%
  \bibfield{title}{%
  \enquote{\bibinfo {title} {Spin-flip excitations, spin waves, and
  magnetoexcitons in graphene landau levels at integer filling factors},}\ }%
  \bibfield{journal}{%
  \bibinfo {journal} {Phys. Rev. B}\ }%
  \textbf{\bibinfo {volume} {82}},\ \bibinfo {pages} {205418} (\bibinfo {year}
  {2010})%
  \bibAnnoteFile{NoStop}{Roldan1}%
\bibitem{Zhang}%
  \BibitemOpen
  \bibfield{author}{%
  \bibinfo {author} {\bibfnamefont{C-H}\ \bibnamefont{Zhang}}\ and\ \bibinfo
  {author} {\bibfnamefont{Y~N}\ \bibnamefont{Joglekar}},\ }%
  \bibfield{title}{%
  \enquote{\bibinfo {title} {Influence of landau-level mixing on wigner
  crystallization in graphene},}\ }%
  \bibfield{journal}{%
  \bibinfo {journal} {Phys. Rev. B}\ }%
  \textbf{\bibinfo {volume} {77}},\ \bibinfo {pages} {205426} (\bibinfo {year}
  {2008})%
  \bibAnnoteFile{NoStop}{Zhang}%
\bibitem{Jung}%
  \BibitemOpen
  \bibfield{author}{%
  \bibinfo {author} {\bibfnamefont{J}~\bibnamefont{Jung}}\ and\ \bibinfo
  {author} {\bibfnamefont{A~H}\ \bibnamefont{MacDonald}},\ }%
  \bibfield{title}{%
  \enquote{\bibinfo {title} {Theory of the magnetic-field-induced insulator in
  neutral graphene sheets},}\ }%
  \bibfield{journal}{%
  \bibinfo {journal} {Phys. Rev. B}\ }%
  \textbf{\bibinfo {volume} {80}},\ \bibinfo {pages} {235417} (\bibinfo {year}
  {2009})%
  \bibAnnoteFile{NoStop}{Jung}%
\bibitem{Kohn}%
  \BibitemOpen
  \bibfield{author}{%
  \bibinfo {author} {\bibfnamefont{W}~\bibnamefont{Kohn}},\ }%
  \bibfield{title}{%
  \enquote{\bibinfo {title} {Cyclotron resonance and de haas-van alphen
  oscillations of an interacting electron gas},}\ }%
  \bibfield{journal}{%
  \bibinfo {journal} {Phys. Rev.}\ }%
  \textbf{\bibinfo {volume} {123}},\ \bibinfo {pages} {1242--1244} (\bibinfo
  {year} {1961})%
  \bibAnnoteFile{NoStop}{Kohn}%
\bibitem{Jiang}%
  \BibitemOpen
  \bibfield{author}{%
  \bibinfo {author} {\bibfnamefont{Z}~\bibnamefont{Jiang}}, \bibinfo {author}
  {\bibfnamefont{E~A}\ \bibnamefont{Henriksen}}, \bibinfo {author}
  {\bibfnamefont{L~C}\ \bibnamefont{Tung}}, \bibinfo {author}
  {\bibfnamefont{Y-J}\ \bibnamefont{Wang}}, \bibinfo {author}
  {\bibfnamefont{M~E}\ \bibnamefont{Schwartz}}, \bibinfo {author}
  {\bibfnamefont{M~Y}\ \bibnamefont{Han}}, \bibinfo {author}
  {\bibfnamefont{P}~\bibnamefont{Kim}},\ and\ \bibinfo {author}
  {\bibfnamefont{H~L}\ \bibnamefont{Stormer}},\ }%
  \bibfield{title}{%
  \enquote{\bibinfo {title} {Infrared spectroscopy of landau levels of
  graphene},}\ }%
  \bibfield{journal}{%
  \bibinfo {journal} {Phys. Rev. Lett.}\ }%
  \textbf{\bibinfo {volume} {98}},\ \bibinfo {pages} {197403} (\bibinfo {year}
  {2007})%
  \bibAnnoteFile{NoStop}{Jiang}%
\bibitem{Henriksen}%
  \BibitemOpen
  \bibfield{author}{%
  \bibinfo {author} {\bibfnamefont{E~A}\ \bibnamefont{Henriksen}}, \bibinfo
  {author} {\bibfnamefont{P}~\bibnamefont{Cadden-Zimansky}}, \bibinfo {author}
  {\bibfnamefont{Z}~\bibnamefont{Jiang}}, \bibinfo {author}
  {\bibfnamefont{Z~Q}\ \bibnamefont{Li}}, \bibinfo {author}
  {\bibfnamefont{L-C}\ \bibnamefont{Tung}}, \bibinfo {author}
  {\bibfnamefont{M~E}\ \bibnamefont{Schwartz}}, \bibinfo {author}
  {\bibfnamefont{M}~\bibnamefont{Takita}}, \bibinfo {author}
  {\bibfnamefont{Y-J}\ \bibnamefont{Wang}}, \bibinfo {author}
  {\bibfnamefont{P}~\bibnamefont{Kim}},\ and\ \bibinfo {author}
  {\bibfnamefont{H~L}\ \bibnamefont{Stormer}},\ }%
  \bibfield{title}{%
  \enquote{\bibinfo {title} {Interaction-induced shift of the cyclotron
  resonance of graphene using infrared spectroscopy},}\ }%
  \bibfield{journal}{%
  \bibinfo {journal} {Phys. Rev. Lett.}\ }%
  \textbf{\bibinfo {volume} {104}},\ \bibinfo {pages} {067404} (\bibinfo {year}
  {2010})%
  \bibAnnoteFile{NoStop}{Henriksen}%
\bibitem{Shizuya2}%
  \BibitemOpen
  \bibfield{author}{%
  \bibinfo {author} {\bibfnamefont{K}~\bibnamefont{Shizuya}},\ }%
  \bibfield{title}{%
  \enquote{\bibinfo {title} {Many-body corrections to cyclotron resonance in
  monolayer and bilayer graphene},}\ }%
  \bibfield{journal}{%
  \bibinfo {journal} {Phys. Rev. B}\ }%
  \textbf{\bibinfo {volume} {81}},\ \bibinfo {pages} {075407} (\bibinfo {year}
  {2010})%
  \bibAnnoteFile{NoStop}{Shizuya2}%
\bibitem{Orlita}%
  \BibitemOpen
  \bibfield{author}{%
  \bibinfo {author} {\bibfnamefont{M}~\bibnamefont{Orlita}}\ and\ \bibinfo
  {author} {\bibfnamefont{M}~\bibnamefont{Potemski}},\ }%
  \bibfield{title}{%
  \enquote{\bibinfo {title} {Dirac electronic states in graphene systems:
  optical spectroscopy studies},}\ }%
  \bibfield{journal}{%
  \bibinfo {journal} {Semicond. Sci. Technol.}\ }%
  \textbf{\bibinfo {volume} {25}},\ \bibinfo {pages} {063001} (\bibinfo {year}
  {2010})%
  \bibAnnoteFile{NoStop}{Orlita}%
\bibitem{Pyatkovskiy}%
  \BibitemOpen
  \bibfield{author}{%
  \bibinfo {author} {\bibfnamefont{P~K}\ \bibnamefont{Pyatkovskiy}}\ and\
  \bibinfo {author} {\bibfnamefont{V~P}\ \bibnamefont{Gusynin}},\ }%
  \bibfield{title}{%
  \enquote{\bibinfo {title} {Dynamical polarization of graphene in a magnetic
  field},}\ }%
  \bibfield{journal}{%
  \bibinfo {journal} {Phys. Rev. B}\ }%
  \textbf{\bibinfo {volume} {83}},\ \bibinfo {pages} {075422} (\bibinfo {year}
  {2011})%
  \bibAnnoteFile{NoStop}{Pyatkovskiy}%
\bibitem{Chiu}%
  \BibitemOpen
  \bibfield{author}{%
  \bibinfo {author} {\bibfnamefont{K~W}\ \bibnamefont{Chiu}}\ and\ \bibinfo
  {author} {\bibfnamefont{J~J}\ \bibnamefont{Quinn}},\ }%
  \bibfield{title}{%
  \enquote{\bibinfo {title} {Plasma oscillations of a two-dimensional electron
  gas in a strong magnetic field},}\ }%
  \bibfield{journal}{%
  \bibinfo {journal} {Phys. Rev. B}\ }%
  \textbf{\bibinfo {volume} {9}},\ \bibinfo {pages} {4724--4732} (\bibinfo
  {year} {1974})%
  \bibAnnoteFile{NoStop}{Chiu}%
\bibitem{Shizuya1}%
  \BibitemOpen
  \bibfield{author}{%
  \bibinfo {author} {\bibfnamefont{K}~\bibnamefont{Shizuya}},\ }%
  \bibfield{title}{%
  \enquote{\bibinfo {title} {Electromagnetic response and effective gauge
  theory of graphene in a magnetic field},}\ }%
  \bibfield{journal}{%
  \bibinfo {journal} {Phys. Rev. B}\ }%
  \textbf{\bibinfo {volume} {75}},\ \bibinfo {pages} {245417} (\bibinfo {year}
  {2007})%
  \bibAnnoteFile{NoStop}{Shizuya1}%
\bibitem{Tahir1}%
  \BibitemOpen
  \bibfield{author}{%
  \bibinfo {author} {\bibfnamefont{M}~\bibnamefont{Tahir}}\ and\ \bibinfo
  {author} {\bibfnamefont{K}~\bibnamefont{Sabeeh}},\ }%
  \bibfield{title}{%
  \enquote{\bibinfo {title} {Inter-band magnetoplasmons in mono- and bilayer
  graphene},}\ }%
  \bibfield{journal}{%
  \bibinfo {journal} {J. Phys.: Condens. Matter}\ }%
  \textbf{\bibinfo {volume} {20}},\ \bibinfo {pages} {425202} (\bibinfo {year}
  {2008})%
  \bibAnnoteFile{NoStop}{Tahir1}%
\bibitem{Berman2}%
  \BibitemOpen
  \bibfield{author}{%
  \bibinfo {author} {\bibfnamefont{O~L}\ \bibnamefont{Berman}}, \bibinfo
  {author} {\bibfnamefont{G}~\bibnamefont{Gumbs}},\ and\ \bibinfo {author}
  {\bibfnamefont{Yu~E}\ \bibnamefont{Lozovik}},\ }%
  \bibfield{title}{%
  \enquote{\bibinfo {title} {Magnetoplasmons in layered graphene structures},}\
  }%
  \bibfield{journal}{%
  \bibinfo {journal} {Phys. Rev. B}\ }%
  \textbf{\bibinfo {volume} {78}},\ \bibinfo {pages} {085401} (\bibinfo {year}
  {2008})%
  \bibAnnoteFile{NoStop}{Berman2}%
\bibitem{Roldan3}%
  \BibitemOpen
  \bibfield{author}{%
  \bibinfo {author} {\bibfnamefont{R}~\bibnamefont{Rold\'{a}n}}, \bibinfo
  {author} {\bibfnamefont{J-N}\ \bibnamefont{Fuchs}},\ and\ \bibinfo {author}
  {\bibfnamefont{M~O}\ \bibnamefont{Goerbig}},\ }%
  \bibfield{title}{%
  \enquote{\bibinfo {title} {Collective modes of doped graphene and a standard
  two-dimensional electron gas in a strong magnetic field: Linear
  magnetoplasmons versus magnetoexcitons},}\ }%
  \bibfield{journal}{%
  \bibinfo {journal} {Phys. Rev. B}\ }%
  \textbf{\bibinfo {volume} {80}},\ \bibinfo {pages} {085408} (\bibinfo {year}
  {2008})%
  \bibAnnoteFile{NoStop}{Roldan3}%
\bibitem{Roldan4}%
  \BibitemOpen
  \bibfield{author}{%
  \bibinfo {author} {\bibfnamefont{R}~\bibnamefont{Rold\'{a}n}}, \bibinfo
  {author} {\bibfnamefont{M~O}\ \bibnamefont{Goerbig}},\ and\ \bibinfo {author}
  {\bibfnamefont{J-N}\ \bibnamefont{Fuchs}},\ }%
  \bibfield{title}{%
  \enquote{\bibinfo {title} {The magnetic field particle-hole excitation
  spectrum in doped graphene and in a standard two-dimensional electron gas},}\
  }%
  \bibfield{journal}{%
  \bibinfo {journal} {Semicond. Sci. Technol.}\ }%
  \textbf{\bibinfo {volume} {25}},\ \bibinfo {pages} {034005} (\bibinfo {year}
  {2010})%
  \bibAnnoteFile{NoStop}{Roldan4}%
\bibitem{Fischer2}%
  \BibitemOpen
  \bibfield{author}{%
  \bibinfo {author} {\bibfnamefont{A~M}\ \bibnamefont{Fischer}}, \bibinfo
  {author} {\bibfnamefont{R~A}\ \bibnamefont{R\"{o}mer}},\ and\ \bibinfo
  {author} {\bibfnamefont{A~B}\ \bibnamefont{Dzyubenko}},\ }%
  \bibfield{title}{%
  \enquote{\bibinfo {title} {Magnetoplasmons and su(4) symmetry in graphene},}\
  }%
  \bibfield{journal}{%
  \bibinfo {journal} {J. Phys.: Conf. Ser.}\ }%
  \textbf{\bibinfo {volume} {286}},\ \bibinfo {pages} {012054} (\bibinfo {year}
  {2011})%
  \bibAnnoteFile{NoStop}{Fischer2}%
\bibitem{Barlas}%
  \BibitemOpen
  \bibfield{author}{%
  \bibinfo {author} {\bibfnamefont{Y}~\bibnamefont{Barlas}}, \bibinfo {author}
  {\bibfnamefont{W-C}\ \bibnamefont{Lee}}, \bibinfo {author}
  {\bibfnamefont{K}~\bibnamefont{Nomura}},\ and\ \bibinfo {author}
  {\bibfnamefont{A~H}\ \bibnamefont{MacDonald}},\ }%
  \bibfield{title}{%
  \enquote{\bibinfo {title} {Renormalized landau levels and particle-hole
  symmetry in graphene},}\ }%
  \bibfield{journal}{%
  \bibinfo {journal} {Int. J. Mod. Phys. B}\ }%
  \textbf{\bibinfo {volume} {23}},\ \bibinfo {pages} {2634--2640} (\bibinfo
  {year} {2009})%
  \bibAnnoteFile{NoStop}{Barlas}%
\bibitem{Gor'kov}%
  \BibitemOpen
  \bibfield{author}{%
  \bibinfo {author} {\bibfnamefont{L~P}\ \bibnamefont{Gor'kov}}\ and\ \bibinfo
  {author} {\bibfnamefont{I~E}\ \bibnamefont{Dzyaloshinskii}},\ }%
  \bibfield{title}{%
  \enquote{\bibinfo {title} {Contribution to the theory of the mott exciton in
  a strong magnetic field},}\ }%
  \bibfield{journal}{%
  \bibinfo {journal} {Sov. Phys. JETP}\ }%
  \textbf{\bibinfo {volume} {26}},\ \bibinfo {pages} {449--453} (\bibinfo
  {year} {1968})%
  \bibAnnoteFile{NoStop}{Gor'kov}%
\bibitem{Ruvinsky}%
  \BibitemOpen
  \bibfield{author}{%
  \bibinfo {author} {\bibfnamefont{Yu~E}\ \bibnamefont{Lozovik}}\ and\ \bibinfo
  {author} {\bibfnamefont{A~M}\ \bibnamefont{Ruvinsky}},\ }%
  \bibfield{title}{%
  \enquote{\bibinfo {title} {Magnetoexcitons in coupled quantum wells},}\ }%
  \bibfield{journal}{%
  \bibinfo {journal} {Phys. Lett. A}\ }%
  \textbf{\bibinfo {volume} {227}},\ \bibinfo {pages} {271--284} (\bibinfo
  {year} {1997})%
  \bibAnnoteFile{NoStop}{Ruvinsky}%
\bibitem{Bolotin}%
  \BibitemOpen
  \bibfield{author}{%
  \bibinfo {author} {\bibfnamefont{K~I}\ \bibnamefont{Bolotin}}, \bibinfo
  {author} {\bibfnamefont{K~J}\ \bibnamefont{Sikes}}, \bibinfo {author}
  {\bibfnamefont{Z}~\bibnamefont{Jiang}}, \bibinfo {author}
  {\bibfnamefont{M}~\bibnamefont{Klima}}, \bibinfo {author}
  {\bibfnamefont{G}~\bibnamefont{Fudenberg}}, \bibinfo {author}
  {\bibfnamefont{J}~\bibnamefont{Hone}}, \bibinfo {author}
  {\bibfnamefont{P}~\bibnamefont{Kim}},\ and\ \bibinfo {author}
  {\bibfnamefont{H~L}\ \bibnamefont{Stormer}},\ }%
  \bibfield{title}{%
  \enquote{\bibinfo {title} {Ultrahigh electron mobility in suspended
  graphene},}\ }%
  \bibfield{journal}{%
  \bibinfo {journal} {Solid State Commun.}\ }%
  \textbf{\bibinfo {volume} {146}},\ \bibinfo {pages} {351--355} (\bibinfo
  {year} {2008})%
  \bibAnnoteFile{NoStop}{Bolotin}%
\bibitem{Ghahari}%
  \BibitemOpen
  \bibfield{author}{%
  \bibinfo {author} {\bibfnamefont{F}~\bibnamefont{Ghahari}}, \bibinfo {author}
  {\bibfnamefont{Y}~\bibnamefont{Zhao}}, \bibinfo {author}
  {\bibfnamefont{P}~\bibnamefont{Cadden-Zimansky}}, \bibinfo {author}
  {\bibfnamefont{K}~\bibnamefont{Bolotin}},\ and\ \bibinfo {author}
  {\bibfnamefont{P}~\bibnamefont{Kim}},\ }%
  \bibfield{title}{%
  \enquote{\bibinfo {title} {Measurement of the nu=1/3 fractional quantum hall
  energy gap in suspended graphene},}\ }%
  \bibfield{journal}{%
  \bibinfo {journal} {Phys. Rev. Lett.}\ }%
  \textbf{\bibinfo {volume} {106}},\ \bibinfo {pages} {046801} (\bibinfo {year}
  {2011})%
  \bibAnnoteFile{NoStop}{Ghahari}%
\bibitem{Elias}%
  \BibitemOpen
  \bibfield{author}{%
  \bibinfo {author} {\bibfnamefont{D~C}\ \bibnamefont{Elias}}, \bibinfo
  {author} {\bibfnamefont{R~V}\ \bibnamefont{Gorbachev}}, \bibinfo {author}
  {\bibfnamefont{A~S}\ \bibnamefont{Mayorov}}, \bibinfo {author}
  {\bibfnamefont{S~V}\ \bibnamefont{Morozov}}, \bibinfo {author}
  {\bibfnamefont{A~A}\ \bibnamefont{Zhukov}}, \bibinfo {author}
  {\bibfnamefont{P}~\bibnamefont{Blake}}, \bibinfo {author}
  {\bibfnamefont{L~A}\ \bibnamefont{Ponomarenko}}, \bibinfo {author}
  {\bibfnamefont{I~V}\ \bibnamefont{Grigorieva}}, \bibinfo {author}
  {\bibfnamefont{K~S}\ \bibnamefont{Novoselov}}, \bibinfo {author}
  {\bibfnamefont{F}~\bibnamefont{Guinea}},\ and\ \bibinfo {author}
  {\bibfnamefont{A~K}\ \bibnamefont{Geim}},\ }%
  \bibfield{title}{%
  \enquote{\bibinfo {title} {Dirac cones reshaped by interaction effects in
  suspended graphene},}\ }%
  \bibfield{journal}{%
  \bibinfo {journal} {Nature Phys.}\ }%
  \textbf{\bibinfo {volume} {7}},\ \bibinfo {pages} {701--704} (\bibinfo {year}
  {2011})%
  \bibAnnoteFile{NoStop}{Elias}%
\bibitem{Knox}%
  \BibitemOpen
  \bibfield{author}{%
  \bibinfo {author} {\bibfnamefont{K~R}\ \bibnamefont{Knox}}, \bibinfo {author}
  {\bibfnamefont{A}~\bibnamefont{Locatelli}}, \bibinfo {author}
  {\bibfnamefont{M~B}\ \bibnamefont{Yilmaz}}, \bibinfo {author}
  {\bibfnamefont{D}~\bibnamefont{Cvetko}}, \bibinfo {author}
  {\bibfnamefont{T~O}\ \bibnamefont{Mentes}}, \bibinfo {author}
  {\bibfnamefont{M~A}\ \bibnamefont{Nino}}, \bibinfo {author}
  {\bibfnamefont{P}~\bibnamefont{Kim}}, \bibinfo {author}
  {\bibfnamefont{A}~\bibnamefont{Morgante}},\ and\ \bibinfo {author}
  {\bibfnamefont{R~M}\ \bibnamefont{Osgood}},\ }%
  \bibfield{title}{%
  \enquote{\bibinfo {title} {Making angle-resolved photoemission measurements
  on corrugated monolayer crystals: Suspended exfoliated single-crystal
  graphene},}\ }%
  \bibfield{journal}{%
  \bibinfo {journal} {Phys. Rev. B}\ }%
  \textbf{\bibinfo {volume} {84}},\ \bibinfo {pages} {115401} (\bibinfo {year}
  {2011})%
  \bibAnnoteFile{NoStop}{Knox}%
\bibitem{Moskalenko1}%
  \BibitemOpen
  \bibfield{author}{%
  \bibinfo {author} {\bibfnamefont{S~A}\ \bibnamefont{Moskalenko}}, \bibinfo
  {author} {\bibfnamefont{M~A}\ \bibnamefont{Liberman}}, \bibinfo {author}
  {\bibfnamefont{P~I}\ \bibnamefont{Khadzhi}}, \bibinfo {author}
  {\bibfnamefont{E~V}\ \bibnamefont{Dumanov}}, \bibinfo {author}
  {\bibfnamefont{Ig~V}\ \bibnamefont{Podlesny}},\ and\ \bibinfo {author}
  {\bibfnamefont{V}~\bibnamefont{Bo\c{t}an}},\ }%
  \bibfield{title}{%
  \enquote{\bibinfo {title} {Influence of excited landau levels on a
  two-dimensional electron-hole system in a strong perpendicular magnetic
  field},}\ }%
  \bibfield{journal}{%
  \bibinfo {journal} {Solid State Commun.}\ }%
  \textbf{\bibinfo {volume} {140}},\ \bibinfo {pages} {236--239} (\bibinfo
  {year} {1996})%
  \bibAnnoteFile{NoStop}{Moskalenko1}%
\bibitem{Moskalenko2}%
  \BibitemOpen
  \bibfield{author}{%
  \bibinfo {author} {\bibfnamefont{S~A}\ \bibnamefont{Moskalenko}}, \bibinfo
  {author} {\bibfnamefont{M~A}\ \bibnamefont{Liberman}}, \bibinfo {author}
  {\bibfnamefont{P~I}\ \bibnamefont{Khadzhi}}, \bibinfo {author}
  {\bibfnamefont{E~V}\ \bibnamefont{Dumanov}}, \bibinfo {author}
  {\bibfnamefont{Ig~V}\ \bibnamefont{Podlesny}},\ and\ \bibinfo {author}
  {\bibfnamefont{V}~\bibnamefont{Bo\c{t}an}},\ }%
  \bibfield{title}{%
  \enquote{\bibinfo {title} {Influence of coulomb scattering of electrons and
  holes between landau levels on energy spectrum and collective properties of
  two-dimensional magnetoexcitons},}\ }%
  \bibfield{journal}{%
  \bibinfo {journal} {Physica E}\ }%
  \textbf{\bibinfo {volume} {39}},\ \bibinfo {pages} {137--149} (\bibinfo
  {year} {2007})%
  \bibAnnoteFile{NoStop}{Moskalenko2}%
\bibitem{Morozov}%
  \BibitemOpen
  \bibfield{author}{%
  \bibinfo {author} {\bibfnamefont{L~A}\ \bibnamefont{Ponomarenko}}, \bibinfo
  {author} {\bibfnamefont{R}~\bibnamefont{Yang}}, \bibinfo {author}
  {\bibfnamefont{T~M}\ \bibnamefont{Mohiuddin}}, \bibinfo {author}
  {\bibfnamefont{M~I}\ \bibnamefont{Katsnelson}}, \bibinfo {author}
  {\bibfnamefont{K~S}\ \bibnamefont{Novoselov}}, \bibinfo {author}
  {\bibfnamefont{S~V}\ \bibnamefont{Morozov}}, \bibinfo {author}
  {\bibfnamefont{A~A}\ \bibnamefont{Zhukov}}, \bibinfo {author}
  {\bibfnamefont{F}~\bibnamefont{Schedin}}, \bibinfo {author}
  {\bibfnamefont{E~W}\ \bibnamefont{Hill}},\ and\ \bibinfo {author}
  {\bibfnamefont{A~K}\ \bibnamefont{Geim}},\ }%
  \bibfield{title}{%
  \enquote{\bibinfo {title} {Effect of a high-$\kappa$ environment on charge
  carrier mobility in graphene},}\ }%
  \bibfield{journal}{%
  \bibinfo {journal} {Phys. Rev. Lett.}\ }%
  \textbf{\bibinfo {volume} {102}},\ \bibinfo {pages} {206603} (\bibinfo {year}
  {2009})%
  \bibAnnoteFile{NoStop}{Morozov}%
\bibitem{Sadowski}%
  \BibitemOpen
  \bibfield{author}{%
  \bibinfo {author} {\bibfnamefont{M~L}\ \bibnamefont{Sadowski}}, \bibinfo
  {author} {\bibfnamefont{G}~\bibnamefont{Martinez}}, \bibinfo {author}
  {\bibfnamefont{M}~\bibnamefont{Potemski}}, \bibinfo {author}
  {\bibfnamefont{C}~\bibnamefont{Berger}},\ and\ \bibinfo {author}
  {\bibfnamefont{W~A}\ \bibnamefont{de~Heer}},\ }%
  \bibfield{title}{%
  \enquote{\bibinfo {title} {Magnetospectroscopy of epitaxial few-layer
  graphene},}\ }%
  \bibfield{journal}{%
  \bibinfo {journal} {Solid State Commun.}\ }%
  \textbf{\bibinfo {volume} {143}},\ \bibinfo {pages} {123--125} (\bibinfo
  {year} {2007})%
  \bibAnnoteFile{NoStop}{Sadowski}%
\bibitem{Kashuba}%
  \BibitemOpen
  \bibfield{author}{%
  \bibinfo {author} {\bibfnamefont{O}~\bibnamefont{Kashuba}}\ and\ \bibinfo
  {author} {\bibfnamefont{V~I}\ \bibnamefont{Fal'ko}},\ }%
  \bibfield{title}{%
  \enquote{\bibinfo {title} {Signature of electronic excitations in the raman
  spectrum of graphene},}\ }%
  \bibfield{journal}{%
  \bibinfo {journal} {Phys. Rev. B}\ }%
  \textbf{\bibinfo {volume} {80}},\ \bibinfo {pages} {241404(R)} (\bibinfo
  {year} {2009})%
  \bibAnnoteFile{NoStop}{Kashuba}%
\bibitem{Giesbers}%
  \BibitemOpen
  \bibfield{author}{%
  \bibinfo {author} {\bibfnamefont{A~J~M}\ \bibnamefont{Giesbers}}, \bibinfo
  {author} {\bibfnamefont{U}~\bibnamefont{Zeitler}}, \bibinfo {author}
  {\bibfnamefont{M~I}\ \bibnamefont{Katsnelson}}, \bibinfo {author}
  {\bibfnamefont{L~A}\ \bibnamefont{Ponomarenko}}, \bibinfo {author}
  {\bibfnamefont{T~M}\ \bibnamefont{Mohiuddin}},\ and\ \bibinfo {author}
  {\bibfnamefont{Maan~J}\ \bibnamefont{C}},\ }%
  \bibfield{title}{%
  \enquote{\bibinfo {title} {Quantum-hall activation gaps in graphene},}\ }%
  \bibfield{journal}{%
  \bibinfo {journal} {Phys. Rev. Lett.}\ }%
  \textbf{\bibinfo {volume} {99}},\ \bibinfo {pages} {206803} (\bibinfo {year}
  {2007})%
  \bibAnnoteFile{NoStop}{Giesbers}%
\bibitem{Berman1}%
  \BibitemOpen
  \bibfield{author}{%
  \bibinfo {author} {\bibfnamefont{O~L}\ \bibnamefont{Berman}}, \bibinfo
  {author} {\bibfnamefont{Yu~E}\ \bibnamefont{Lozovik}},\ and\ \bibinfo
  {author} {\bibfnamefont{G}~\bibnamefont{Gumbs}},\ }%
  \bibfield{title}{%
  \enquote{\bibinfo {title} {Bose-einstein condensation and superfluidity of
  magnetoexcitons in bilayer graphene},}\ }%
  \bibfield{journal}{%
  \bibinfo {journal} {Phys. Rev. B}\ }%
  \textbf{\bibinfo {volume} {77}},\ \bibinfo {pages} {155433} (\bibinfo {year}
  {2008})%
  \bibAnnoteFile{NoStop}{Berman1}%
\bibitem{Fil}%
  \BibitemOpen
  \bibfield{author}{%
  \bibinfo {author} {\bibfnamefont{D~V}\ \bibnamefont{Fil}}\ and\ \bibinfo
  {author} {\bibfnamefont{Kravchenko~L}\ \bibnamefont{Yu}},\ }%
  \bibfield{title}{%
  \enquote{\bibinfo {title} {Superfluid state of magnetoexcitons in double
  layer graphene structures},}\ }%
  \bibfield{journal}{%
  \bibinfo {journal} {AIP Conf. Proc.}\ }%
  \textbf{\bibinfo {volume} {1198}},\ \bibinfo {pages} {34--41} (\bibinfo
  {year} {2009})%
  \bibAnnoteFile{NoStop}{Fil}%
\bibitem{Bezuglyi}%
  \BibitemOpen
  \bibfield{author}{%
  \bibinfo {author} {\bibfnamefont{A~I}\ \bibnamefont{Bezugly\v{i}}},\ }%
  \bibfield{title}{%
  \enquote{\bibinfo {title} {Dynamical equation for an electron-hole pair
  condensate in a system of two graphene layers},}\ }%
  \bibfield{journal}{%
  \bibinfo {journal} {Low Temp. Phys.}\ }%
  \textbf{\bibinfo {volume} {36}},\ \bibinfo {pages} {236--242} (\bibinfo
  {year} {2010})%
  \bibAnnoteFile{NoStop}{Bezuglyi}%
\bibitem{Berman3}%
  \BibitemOpen
  \bibfield{author}{%
  \bibinfo {author} {\bibfnamefont{O~L}\ \bibnamefont{Berman}}, \bibinfo
  {author} {\bibfnamefont{R~Ya}\ \bibnamefont{Kezerashvili}},\ and\ \bibinfo
  {author} {\bibfnamefont{Yu~E}\ \bibnamefont{Lozovik}},\ }%
  \bibfield{title}{%
  \enquote{\bibinfo {title} {Bose-einstein condensation of trapped polaritons
  in two-dimensional electron-hole systems in a high magnetic field},}\ }%
  \bibfield{journal}{%
  \bibinfo {journal} {Phys. Rev. B}\ }%
  \textbf{\bibinfo {volume} {80}},\ \bibinfo {pages} {115302} (\bibinfo {year}
  {2009})%
  \bibAnnoteFile{NoStop}{Berman3}%
\bibitem{Fischer1}%
  \BibitemOpen
  \bibfield{author}{%
  \bibinfo {author} {\bibfnamefont{A~M}\ \bibnamefont{Fischer}}, \bibinfo
  {author} {\bibfnamefont{R~A}\ \bibnamefont{R\"{o}mer}},\ and\ \bibinfo
  {author} {\bibfnamefont{A~B}\ \bibnamefont{Dzyubenko}},\ }%
  \bibfield{title}{%
  \enquote{\bibinfo {title} {Symmetry content and spectral properties of
  charged collective excitations for graphene in strong magnetic fields},}\ }%
  \bibfield{journal}{%
  \bibinfo {journal} {Europhys. Lett.}\ }%
  \textbf{\bibinfo {volume} {92}},\ \bibinfo {pages} {37003} (\bibinfo {year}
  {2010})%
  \bibAnnoteFile{NoStop}{Fischer1}%
\bibitem{Roldan2}%
  \BibitemOpen
  \bibfield{author}{%
  \bibinfo {author} {\bibfnamefont{R}~\bibnamefont{Rold\'{a}n}}, \bibinfo
  {author} {\bibfnamefont{M~O}\ \bibnamefont{Goerbig}},\ and\ \bibinfo {author}
  {\bibfnamefont{J-N}\ \bibnamefont{Fuchs}},\ }%
  \bibfield{title}{%
  \enquote{\bibinfo {title} {Theory of bernstein modes in graphene},}\ }%
  \bibfield{journal}{%
  \bibinfo {journal} {Phys. Rev. B}\ }%
  \textbf{\bibinfo {volume} {83}},\ \bibinfo {pages} {205406} (\bibinfo {year}
  {2011})%
  \bibAnnoteFile{NoStop}{Roldan2}%
\bibitem{Goerbig1}%
  \BibitemOpen
  \bibfield{author}{%
  \bibinfo {author} {\bibfnamefont{M~O}\ \bibnamefont{Goerbig}}, \bibinfo
  {author} {\bibfnamefont{J-N}\ \bibnamefont{Fuchs}}, \bibinfo {author}
  {\bibfnamefont{K}~\bibnamefont{Kechedzhi}},\ and\ \bibinfo {author}
  {\bibfnamefont{V~I}\ \bibnamefont{Fal'ko}},\ }%
  \bibfield{title}{%
  \enquote{\bibinfo {title} {Filling-factor-dependent magnetophonon resonance
  in graphene},}\ }%
  \bibfield{journal}{%
  \bibinfo {journal} {Phys. Rev. Lett.}\ }%
  \textbf{\bibinfo {volume} {99}},\ \bibinfo {pages} {087402} (\bibinfo {year}
  {2007})%
  \bibAnnoteFile{NoStop}{Goerbig1}%
\end{thebibliography}%

\end{document}